\documentclass[journal]{IEEEtran}

\usepackage{amssymb, amsmath, amsthm}
\usepackage{amsfonts, bbm}
\usepackage{graphicx}
\usepackage{epstopdf}
\usepackage{times}
\usepackage{float}
\usepackage{array}
\usepackage{arydshln}
\usepackage{bibentry}
\usepackage{booktabs}
\usepackage{cite}
\usepackage[caption=false,font=footnotesize]{subfig}
\usepackage{color}

\newtheorem{thm}{\itshape \indent Theorem}
\newtheorem{lem}{\itshape \indent Lemma}
\newtheorem{col}{\itshape \indent Corollary}
\theoremstyle{remark}

\DeclareMathOperator*{\Exp}{Exp}
\newcommand*{\Resize}[2]{\resizebox{#1}{!}{$#2$}}%
\newcommand*{\Scale}[2][4]{\scalebox{#1}{$#2$}}%

\begin{document}

\title{Location-Aware Cross-Tier Coordinated Multipoint Transmission in Two-Tier Cellular Networks}

\author{Ahmed Hamdi Sakr and Ekram Hossain
\thanks{A. H. Sakr and E. Hossain are with the Department of Electrical and Computer Engineering, University of Manitoba, Winnipeg, Canada (emails:  Ahmed.Sakr@umanitoba.ca, Ekram.Hossain@umanitoba.ca). This work was supported  by a Strategic Project Grant (STPGP 430285) from the Natural Sciences and Engineering Research Council of Canada (NSERC).}
}

\maketitle

\begin{abstract}
Multi-tier cellular networks are considered as an effective solution to enhance the coverage and data rate offered by cellular systems. In a multi-tier network, high power base stations (BSs) such as macro BSs are overlaid by lower power small cells such as femtocells and/or picocells. However, co-channel deployment of  multiple tiers of BSs gives rise to the problem of cross-tier interference that significantly impacts the performance of wireless networks. Multicell cooperation techniques, such as coordinated multipoint (CoMP) transmission,  {have} been proposed as a promising solution to mitigate the impact of the cross-tier interference in multi-tier networks. In this paper, we propose a novel scheme for \textbf{L}ocation-\textbf{A}ware \textbf{C}ross-\textbf{T}ier \textbf{C}ooperation (LA-CTC) between BSs in different tiers for downlink CoMP transmission in two-tier cellular networks. On one hand, the proposed scheme only uses CoMP transmission to enhance the performance of the users who suffer from high cross-tier interference due to the co-channel deployment of small cells such as picocells. On the other hand, users with good signal-to-interference-plus-noise ratio (${\rm SINR}$) conditions are served directly by a single BS from any of the two tiers. Thus, the data exchange between the cooperating BSs over the backhaul network can be reduced when compared to the traditional CoMP transmission scheme. We use tools from stochastic geometry to quantify the performance gains obtained by using the proposed scheme in terms of outage probability, achievable data rate, and load per BS. We compare the performance of the proposed scheme with that of other schemes in the literature such as the schemes which use cooperation to serve all users and schemes that use range expansion to offload users to the small cell tier. 
\end{abstract}


\section{Introduction}
Deployment of multi-tier networks is an attractive solution to satisfy the ever-increasing users' demand for higher data rates and network coverage. Unlike traditional single-tier networks, multi-tier networks consist of different classes of base stations (BSs) such as  femto base stations and pico base stations. These BSs operate simultaneously in the same geographical area and differ in transmit power, coverage range, and spatial density \cite{damnjanovic2011survey}. However, with co-channel deployment of multiple network tiers, cross-tier interference degrades  network performance in terms of coverage and throughput. For example, macro users  located in the close vicinity of a small cell may be victimized by transmissions to small cell users. The concept of cooperation has been proposed as one solution to address the interference problem \cite{gesbert2010multi, simeone2009local}. For example, coordinated multipoint (CoMP) transmission (also referred as network MIMO) is one form of cooperation in which multiple BSs communicate with each other to cancel out the interference and improve the overall system performance by jointly  {transmitting} the users' data concurrently \cite{lee2012coordinated, venkatesan2007network, foschini2006coordinating, irmer2011coordinated}. In CoMP, BSs use backhaul links to exchange users' data and/or control information where these links are capacity-limited in practice and affect the performance of the wireless system  {\cite{marsch2008base}}. 

\begin{figure}[!t]
\centering
\includegraphics[width=0.48\textwidth]{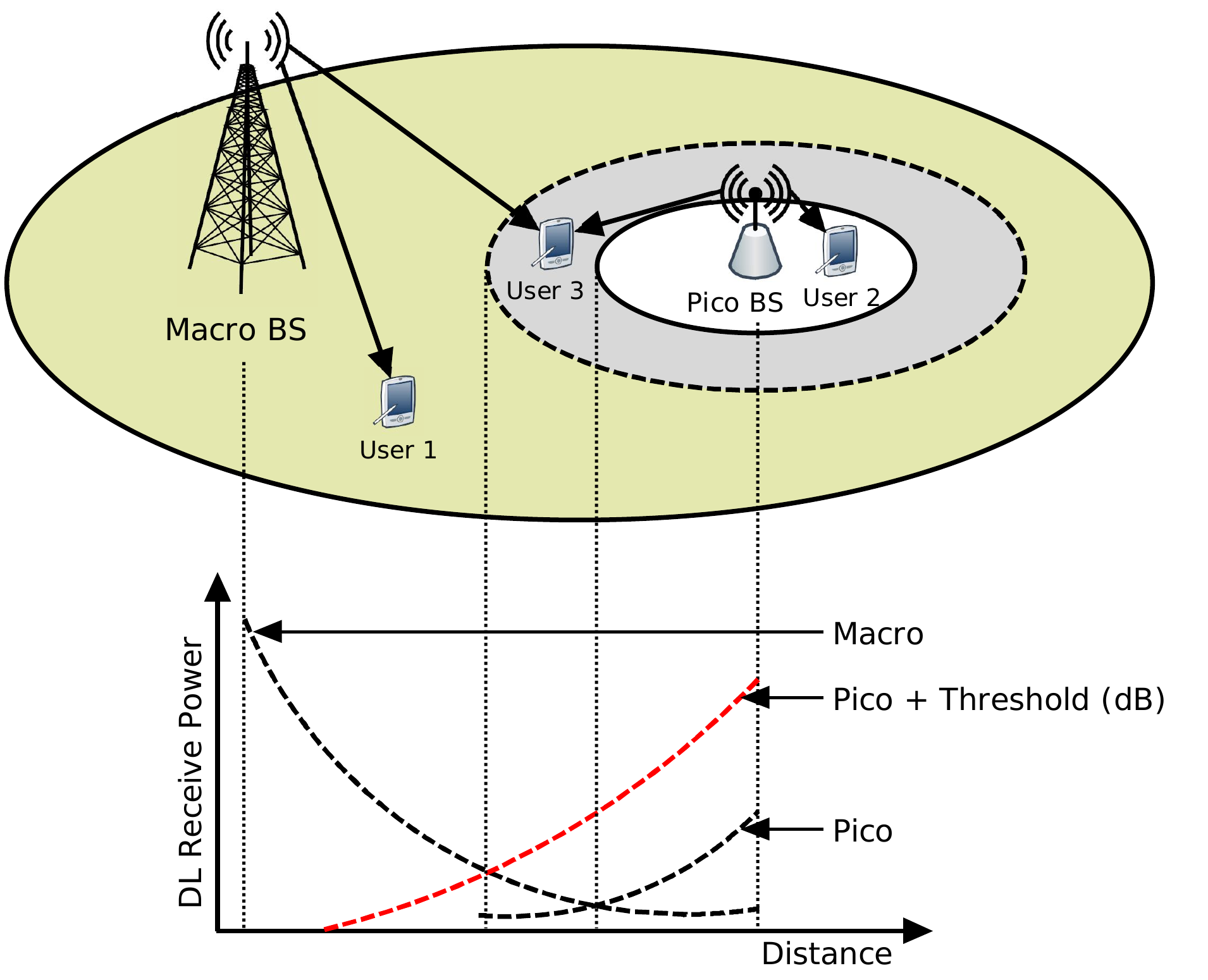}
\caption{A two-tier cellular network with a macrocell and a picocell where the range of cooperation is defined by a positive threshold. While each of User 1 and User 2 is served by only one BS that results in the maximum received power from any of the two tiers, User 3 is connected to more than one BS -- one BS from each tier that results in the maximum received power from that tier.}
\label{sysModel}
\end{figure} 

Multicell cooperation solutions such as CoMP could be effective to mitigate the effect of cross-tier interference in multi-tier networks. For example,  in the two-tier macrocell-picocell network shown in Fig. \ref{sysModel}, although the power received at User $3$ from the serving macro BS is higher than that of the interference resulting from the closest pico BS, the interference power from the closest pico BS can be comparable to the useful signal power which results in a low value of signal-to-interference-plus-noise ratio (${\rm SINR}$). Therefore, the macro BS can cooperate with the interfering pico BS to serve User 3 jointly. This will not only eliminate the strongest interference signal, but also increase the  useful received signal  power  by taking the advantage of the user's proximity to that interferer thus improving the ${\rm SINR}$. However, using cooperation might be unnecessary in some cases. For example, in Fig. \ref{sysModel}, the useful signal power received at User $1$ and User $2$ from the serving macro BS and pico BS, respectively, is sufficiently higher than the power received from the strongest interferer, i.e., the pico BS for User $1$ and the macro BS for User $2$. For these two users, the gain of cooperation may not be high compared to the costs of joint processing and using the backhaul network to exchange users' data especially when the  capacity of the backhaul links is limited.

In this paper, to improve coverage and throughput in a two-tier macrocell-picocell network, we propose a novel location-aware cross-tier cooperation (LA-CTC) scheme in which macro BSs and small cells can cooperate to serve a user jointly only if the user suffers from high interference due to the deployment of pico cells. This user is then referred to be served in {\em CoMP mode}. Otherwise, if the power received from the interfering BS at the user is not high enough to cause severe interference, direct link transmission is used to serve the user without cooperation and the user is referred to be served in {\em non-CoMP mode}. Note that the main focus of this work is on mitigating the effect of cross-tier interference.

As shown in Fig. \ref{sysModel}, we define a region around the picocell in which a user is served by CoMP transmission; otherwise, the user is connected directly to one BS, i.e., macro BS or pico BS. That is, when the ratio of the received power from the macro BS at any user to the received power from the pico BS exceeds a predefined threshold (greater than $1$)\footnote{This threshold is referred to as the {\em cooperation threshold}, which will be defined later in Section III.B.}, this implies that the useful signal is sufficiently higher than the interference, and thus, cooperation is unnecessary and the user (e.g. User $1$) is served by the macro BS only. On the other hand, if the ratio is less than the predefined threshold and still greater than $1$, cooperation is beneficial since the interference power is comparable to the useful signal power (e.g., for User $3$). Finally, if the ratio is less than $1$, the user (e.g., User $2$) is directly connected to the small cell since the small cell is stronger than the macro cell in this case.  

The main motivation of the proposed scheme is to provide better coverage in multi-tier networks while considering the limitation of the backhaul network. For example, assume a macrocell-picocell network where  each macrocell has $p$ randomly-located pico BSs within its coverage area. Since cooperation in the LA-CTC scheme is only possible between BSs belonging to different tiers, only $p$ backhaul links per macrocell are required to enable cooperation between a macro BS and pico BSs in its coverage (a star-connected backhaul network). Now consider another scheme where cooperation is also allowed between BSs belonging to the same tier. In this case both macro and pico BSs are required to exchange user's data in order to perform joint transmission (a fully-connected mesh backhaul network). For example, when cooperation is limited between pico BSs within the same macrocell, this needs $\binom{p}{2}$ backhaul links to connect any two pico BSs. In addition, each macro BS should be able to exchange users' data with at least its first $q$ neighbors as well as the $p$ pico BSs in its coverage. In total, at least $0.5 p^2 + 0.5 p + q$ backhaul links are required to enable cooperation between the BSs. Although the latter scheme offers a better coverage when compared to the LA-CTC scheme,  only $p$ backhaul links are needed for the LA-CTC scheme. Therefore, with the LA-CTC scheme, there is a significant  saving in the number of backhaul links when the number of pico BSs per macrocell is large.

 Note that other techniques, such as range expansion (also referred as flexible cell association), have also been proposed to improve the performance of multi-tier networks and balance the load for all tiers. For example, in a two-tier network with range expansion, users from the macro-tier are offloaded to the small cell tier, where the association to the small cells is biased. That is, a positive bias factor is added to the power of the pilot signals of the small cell base stations to convince the macro users who are close to a small cell coverage boundary to connect to that small cell even if the power received from the macro BS  is stronger than that received from the small cell base station \cite{okino2011pico, lopez2011enhanced, 6287527}. 

We analyze the performance of the proposed LA-CTC scheme for downlink transmission in a two-tier cellular network. We use tools from stochastic geometry to model the network where the locations of the BSs in each tier are distributed according to a two-dimensional independent homogeneous Poisson Point Process (PPP)  {\cite{dhillon2012modeling}}. Each tier of BSs is characterized by its available transmit power, intensity, and path-loss exponent value. In order to evaluate the performance of the proposed scheme, we derive closed-form expressions for the outage probability and data rate. Furthermore, we use our analytical model to derive expressions for outage probability for the range expansion scheme, the scheme with full cooperation where all users are served by cross-tier coordinated CoMP transmission, as well as the traditional scheme where neither cooperation nor range expansion is used. The performances of the different schemes are compared in terms of outage probability, average achievable data rate, and load per BS. The results show that the proposed cooperation scheme outperforms the traditional  range expansion scheme for multi-tier networks in terms of both outage and data rate, while it has higher load per BS. Compared to the full cooperation scheme, the proposed scheme reduces the amount of users' data exchange over the backhaul network as measured by the load per BS. In addition, the outage performance of the proposed scheme approaches that with full cooperation for a  wide range of values of cooperation threshold.

The contributions of the paper can be summarized as follows:
\begin{itemize}
	\item We propose a novel user-centric location-aware cross-tier cooperation (LA-CTC) scheme that uses CoMP transmission for users who experience high levels of interference power compared to the power level of the useful signal received from the serving BS. We define a range of interference power based on which the transmission mode (i.e., CoMP or non-CoMP transmission) is decided by each user individually. 
	\item We use stochastic geometry to evaluate the performance of the proposed scheme in terms of the outage probability, average rate, and load per BS as our key metrics. We compare the proposed scheme with other schemes  such as the range expansion scheme, full cooperation scheme, as well as a non-cooperative scheme in which a user is served by the strongest BS only. The derived expressions are in the closed-integral form.
	\item We analyze the performance of the different schemes under different system parameters by varying the BS intensities, path-loss exponents, cooperation range, and required ${\rm SINR}$ thresholds. Then, we highlight the insights obtained from the analysis and show the impact of the aforementioned parameters on the network behavior.
	\item We show that the proposed LA-CTC scheme is promising for improving the network outage and achievable spectral efficiency while considering the load of macro BSs. Furthermore, we show that the performance of the LA-CTC scheme lies in the middle between the  {performance} of the traditional range expansion-based networks and full cooperation networks. 
\end{itemize}

The rest of this paper is organized as follows. A review of the related work is presented in Section~\ref{sec:related}. The system model, different modes of operation of the users, probability of a user to operate in a certain mode, as well as the distance analysis for the users in different modes are presented in Section \ref{sec:sysModel}. In Section \ref{sec:out_rate}, the outage probability and ergodic rate are obtained for the users in different modes. Finally, the performance evaluation results are presented in Section \ref{sec:results} and the paper is concluded in Section~\ref{sec:conc}.

\section{Related Work}
\label{sec:related}
Previous works on multi-tier networks and multicell cooperation can be divided into two general groups. In the first group, statistical modeling techniques, such as stochastic geometry, are used to analyze network performance and obtain statistically-optimal decision parameters \cite{dhillon2012modeling, 6287527, martin2013comp, giovanidis2013stochastic, 6410048, xia2012downlink,bennis1}. In the second group, instantaneous optimal decisions are obtained by using the instantaneous information of the network based on some objective function \cite{116565, sd5f165, sakr2014icc, sakr2014icc2}. Note that the statistically-optimal parameters might not be optimal on a short time-scale, however, obtaining instantaneous optimal parameters costs more signaling and computations.  

 In \cite{dhillon2012modeling} the authors provide a general framework to analyze and evaluate the performance of a cellular network with $K$ tiers of BSs. In this model, independent PPPs are used to capture the randomness of the locations of BSs as well as the differences in transmit power, propagation environment, and BS spatial density. In addition, analytical expressions for outage probability, achievable data rate, and load per BS are obtained. In \cite{6287527}, the model is extended where the association to different tiers is biased (range expansion). It shows that range expansion degrades the overall network performance in terms of outage and rate. On the other hand, in the context of multicell cooperation, the authors in \cite{martin2013comp} propose two clustering schemes for CoMP transmission in multi-tier networks where clustering is performed on a per-user basis and the performance is evaluated in terms of outage probability. It is assumed that the backhaul network is ideal and the number of cooperating BSs in each cluster is constant. While the first clustering scheme forms a group of $N$ BSs which results in the reception of the $N$ strongest signals at the receiver, the second clustering scheme selects the $N$ closest BSs to the receiver where one BS is chosen from each tier.

In \cite{giovanidis2013stochastic}, the authors propose a cooperation scheme to mitigate the co-tier interference for single-tier networks in which a user-centric decision criterion is used to decide whether to be served with or without cooperation.  The decision is based on the distance between the user and its first two neighboring BSs and some decision parameters. All BSs are assumed to be able to exchange users' data to perform joint transmission with power splitting. The authors use stochastic geometry to investigate the effect of limited channel state information (CSI) at the transmitter. The authors in \cite{6410048} propose another clustering scheme for single-tier networks  where the clusters are formed in a random manner by grouping the BSs that lie in the same Voronoi cell of an overlaying PPP with low intensity. In this paper, BSs that belong to the same cluster cooperate to nullify the interference by exchanging the CSI data. In \cite{xia2012downlink}, the authors use stochastic geometry to evaluate the impact of the overhead delay on the performance of CoMP transmission in multi-tier networks where with zero-forcing beamforming (ZFBF) as a precoding scheme. In \cite {bennis1}, a macrocell-femtocell network with single macro user and macro BS is considered where all femto BSs are cognitive. To mitigate the cross-tier interference, the macro user is assumed to generate a busy tone such that femto BSs defer their transmissions if the received power is greater than a predefined threshold. The authors use stochastic geometry to obtain the outage probability and average data rate.

 The authors in \cite{116565} derive closed-form expressions for the bias factor of range expansion in a picocell-macrocell network for downlink and uplink. Furthermore, a cooperative scheduling scheme between macro and pico BSs is proposed to mitigate the effect of high interference in the expanded regions where simulations are used to evaluate the network and the proposed scheme. In \cite{sd5f165}, a game-theoretic approach is used to study the impact of the backhaul constraints on the performance of femtocell networks with CoMP transmissions. A cooperative game is formulated such that each femtocell chooses the cooperation strategy and exchanges users' data to its cooperative partner over either wired or wireless backhaul. The objective of the proposed game is to balance the tradeoff between the achievable spectral efficiency and delay. The authors in \cite{sakr2014icc2} and \cite{sakr2014icc} use fractional programming to obtain the optimal power, channel, and precoding coefficients allocation for CoMP transmissions in single-tier and two-tier cellular networks, respectively. In both works, the optimization problem aims at maximizing the energy efficiency (bit/Joule) under co-tier or cross-tier interference, power budget, and backhaul link capacity constraints.

To the best of our knowledge, the concepts of cross-tier BS cooperation along with location-aware BS cooperation, which are introduced in this paper, have not been explored previously in the literature\footnote{We use the term ``location-aware'' in the sense that the locations of both users and BSs are considered to make a decision on whether cooperative transmission will be used or not.}. Note that, we have proposed a similar location-aware BS cooperation scheme for single-tier cellular networks in \cite{sakr2014icc2} to mitigate the effect of the co-tier interference.

\section{System Model and Assumptions}
\label{sec:sysModel}
\subsection{Two-tier Cellular Network Model}

We consider downlink transmission in a two-tier macrocell-picocell network where both tiers are independent with different spatial densities, path-loss exponents, and transmit powers. BSs belonging to the same tier $i\in \{ 1,2\}$ have the same transmit power $P_i$. Locations of BSs in the $i^{th}$ tier are modeled according to a two-dimensional homogeneous PPP $\Phi_i \in \mathbb{R}^2$ with spatial intensity $\lambda_i$.  {The users are spatially distributed according to some independent stationary point process $\Phi_u \in \mathbb{R}^2$ (e.g., a homogenous PPP) with intensity $\lambda_u$ which is assumed high enough (compared to $\lambda_1$) such that each BS has at least one user to serve. For statistical analysis, without any loss of generality, we consider a typical user at the origin  \cite{baccelli2010stochastic}. During a transmission interval, a user served by a macro BS and/or a pico BS in a particular channel will experience interference from the other macro BSs and pico BSs. However, there will be no intra-cell interference assuming that different users in a cell are served using orthogonal time-frequency resources (e.g. OFDMA). Different macrocells and picocells can use the same channels (i.e., a co-channel deployment scenario is considered).} All transmitters and receivers are equipped with a single antenna. 

\begin{figure}[!t]
\centering
\includegraphics[width=0.47\textwidth]{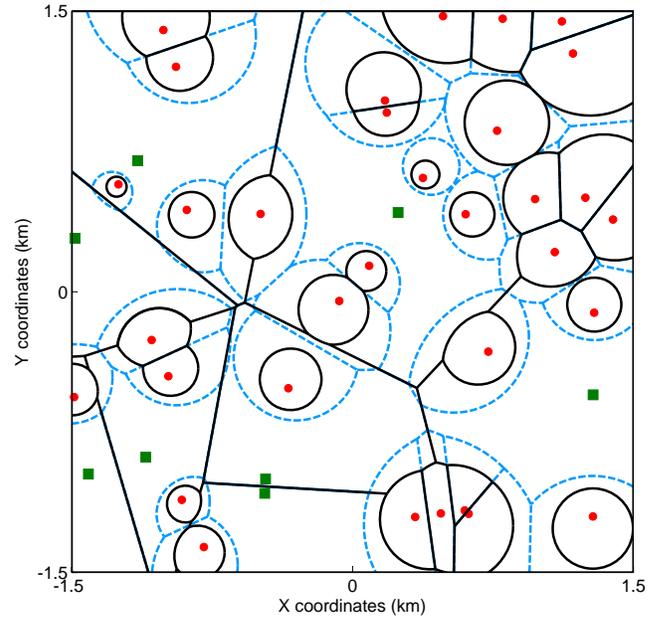}
\caption{A two-tier cellular network with a macro-tier (squares) overlaid with lower power and denser picocells (red circles). Solid black lines show the coverage area of each cell for a traditional two-tier network, while the dashed lines show the cooperation regions that surround each picocell in which cooperation is performed between the macro and pico tiers.}
\label{map}
\end{figure}

Without loss of generality, Fig. \ref{map} shows a realization of a two-tier cellular network where a macrocell network tier (or macro-tier) is deployed as tier $1$ and overlaid with a denser and lower power picocell network tier (or pico-tier) as tier $2$. For a generic point $y \in \mathbb{R}^2$, we define $x_i$ as the BS belonging to the $i^{th}$ tier that results in the strongest long-term average received power at this point. That is,
\begin{align}
x_i &= \arg \max_{x \in \Phi_i} \{ P_i \|x - y\|^{-\alpha_i} \} \label{eq:xs}
\end{align}
where different path-loss exponents $\{\alpha_i\}_{i=1,2}$ are used for downlink modeling at each network tier and $\|\cdot\|$ denotes the Euclidean distance. 
 
\subsection{Mode of Operation and User Association: Location-Aware Cross-tier Cooperation}

 {Based on the received power from each tier, each user independently chooses its mode of operation through or without cooperation.} In this context, we define two modes of operation: non-CoMP and CoMP transmission modes. In the {\em non-CoMP mode} of operation, the user is connected to the BS that results in the maximum long-term average received power regardless of the corresponding network tier, i.e., macro-tier or pico-tier. In the {\em CoMP mode} of operation, the user is served by two BSs that cooperate with each other to jointly transmit data to this user. In this mode, one BS is selected from each tier based on the maximum received power at the user. That is, the users are split into three disjoint groups: non-CoMP macro users, non-CoMP pico users, and CoMP users.  

To elaborate, if the received signal power from the strongest BS at the user is sufficiently higher than that received from the highest interferer, the user operates in the non-CoMP mode since the cooperation between the serving BS and this interfering BS is not necessary in this case. On the other hand, if the received signal power at the user from the strongest interfering BS is comparable to the useful signal power received from the strongest BS, the user operates in the CoMP mode. In this case, the network takes advantage of the proximity of the interfering BS to the user and makes it to cooperate with the user's serving BS to jointly transmit data to the user. This not only mitigates the effect of the highest interferer, but also increases the power level of the useful signal.

 We define $\mathcal{B}$ as the set of BSs that serve a typical user, which can be written as follows:
\begin{align}
\mathcal{B} = \left\{
\begin{array}{l l l}
\{x_1\}, & \text{if } \frac{P_1 R_1^{-\alpha_1}}{P_2 R_2^{-\alpha_2}} \geq \beta & \text{ {``non-CoMP macro''}}\\
\{x_2\}, & \text{if } \frac{P_1 R_1^{-\alpha_1}}{P_2 R_2^{-\alpha_2}} \leq 1& \text{ {``non-CoMP pico''}}\\
\{x_1, x_2\}, & \text{if } 1 < \frac{P_1 R_1^{-\alpha_1}}{P_2 R_2^{-\alpha_2}} < \beta& \text{ {``CoMP''}}
\end{array}\right. \label{eq:coop_set}
\end{align}
where $x_i$ ($i \in \{1,2 \}$) is defined in (\ref{eq:xs}) and $R_i$ ($i \in \{1,2 \}$) is the distance from the typical user to the strongest BS in the $i^{th}$ tier.  $\beta$ is cooperation threshold which represents the ratio between the powers received from the serving macro BS and the strongest pico BS, respectively. This threshold defines the level of cross-tier interference beyond which the user switches to the CoMP mode. That is, if the strongest interference power received from the pico-tier $P_{int}$ at some macro user is in the range $\frac{1}{\beta} P_1 R_1^{-\alpha_1}<P_{int}<P_1 R_1^{-\alpha_1}$, this user switches its mode to be served via cooperation. 

As shown in Fig. \ref{sysModel}, User 1's received power from the macro BS is stronger than that from the pico BS plus the threshold (dB) and User 2's received power from the pico BS is stronger than that from the macro BS. Therefore, User 1 and 2 operate in the non-CoMP mode where they are associated with the macro BS and pico BS, respectively. On the other hand, although User 3's received power from the macro BS is higher than that from the pico BS, the received power from the pico BS plus the threshold (dB) is higher. Therefore, User 3 operates in the CoMP transmission mode where it is connected to both the BSs.

For the proposed LA-CTC scheme, cooperation threshold $\beta$ is an important design parameter and plays a key role in controlling the gains obtained by using this scheme. That is, the higher the cooperation threshold, the larger is the cooperation region which improves the overall system performance while increasing the amount of data exchange over the backhaul network as well as the load per BS. On the other hand, the lower the cooperation threshold, the smaller is the cooperation region which reduces the backhaul signaling between BSs and the load per BS while sacrificing some overall performance gain in terms of outage and data rate.

\subsection{Distance Analysis}

Let $q_M$, $q_P$, and $q_C$ denote the probability that a typical user is in non-CoMP mode and served by the macro BS (i.e., non-CoMP macro user), in the non-CoMP mode and served by the pico BS (i.e., non-CoMP pico user), and in the CoMP mode, respectively.  {Conditioned on each event, in the following lemma, we derive the probability density functions (PDFs) of the distance between a typical user at the origin and its serving BS(s) in the different modes of operation. 

For a typical user in the CoMP mode, we denote by $f_{R_C}(\mathbf{r})$ the joint PDF of the distances between the typical user and its two serving BSs $x_1$ and $x_2$, i.e., macro BS and pico BS, respectively. For a non-CoMP macro user, we denote by $f_{R_1}(r) $ the PDF of the distance between a macro user and its serving macro BS $x_1$. Finally, $f_{R_2}(r) $ is the PDF of the distance between a non-CoMP pico user and its serving pico BS $x_2$.}

\begin{lem}\normalfont 
The PDFs of the distances between a typical user and its serving BS(s) are
\begin{align}
&f_{R_1}(r) = \frac{2\pi \lambda_1}{q_M} r \exp \Resize{4.65cm}{\left[ -\pi \left( \lambda_1 r^2 + \lambda_2 \left(\frac{\beta P_2}{P_1}\right)^{\frac{2}{\alpha_2}} r^{\frac{2\alpha_1}{\alpha_2}} \right) \right]} \label{eq:dist_M}\\
&f_{R_2}(r) = \frac{2\pi \lambda_2}{q_P} r \exp \Resize{4.65cm}{\left[ -\pi \left( \lambda_2 r^2 + \lambda_1 \left(\frac{P_1}{P_2}\right)^{\frac{2}{\alpha_1}} r^{\frac{2\alpha_2}{\alpha_1}} \right) \right]},\label{eq:dist_P}\\
&f_{R_C}(\mathbf{r}) = \frac{4 \pi^2 \lambda_1 \lambda_2}{q_C} r_1 r_2 \exp \left[ -\pi \left( \lambda_1 r_1^2 + \lambda_2 r_2^2 \right) \right] \label{eq:dist_C}
\end{align}
where $r_1 \in \mathbb{R}^+$ and $\left(\frac{P_2}{P_1}\right)^{\frac{1}{\alpha_2}} r_1^{\frac{\alpha_1}{\alpha_2}} < r_2 < \left(\frac{ \beta P_2}{P_1}\right)^{\frac{1}{\alpha_2}} r_1^{\frac{\alpha_1}{\alpha_2}}$, and
\begin{align}
q_M &= 2\pi \lambda_1 \int \limits_{\mathbb{R}_+}\!  r \exp \Resize{4.7cm}{\left[ -\pi \left( \lambda_1 r^2 + \lambda_2 \left(\frac{\beta P_2}{P_1}\right)^{\frac{2}{\alpha_2}} r^{\frac{2\alpha_1}{\alpha_2}} \right) \right]} \text{d}r \label{eq:q_M}\\
q_P &= 2\pi \lambda_2 \int \limits_{\mathbb{R}_+}\! r \exp \Resize{4.7cm}{\left[ -\pi \left( \lambda_2 r^2 + \lambda_1 \left(\frac{P_1}{P_2}\right)^{\frac{2}{\alpha_1}} r^{\frac{2\alpha_2}{\alpha_1}} \right) \right]} \text{d}r, \label{eq:q_P}\\
q_C &= 1 - q_M - q_P. \label{eq:q_C}
\end{align}
\label{lem1}
\begin{IEEEproof}
See \textbf{Appendix \ref{lem1proof}}.
\end{IEEEproof}
\end{lem}

For the special case when $\alpha_1=\alpha_2=\alpha$, $q_M$ and $q_P$ can be expressed in a closed-form as
\begin{align}
q_M = \frac{\lambda_1 P_1^{\frac{2}{\alpha}}}{\lambda_1 P_1^{\frac{2}{\alpha}} + \lambda_2 (\beta P_2)^{\frac{2}{\alpha}}},\quad q_P = \frac{\lambda_2 P_2^{\frac{2}{\alpha}}}{\lambda_1 P_1^{\frac{2}{\alpha}} + \lambda_2 P_2^{\frac{2}{\alpha}}}. \label{eq:q_same}
\end{align}
Furthermore, it can be seen that when the cooperation threshold $\beta$ is set to $1$ (no cooperation), the probability that a typical user operates in the CoMP mode reduces to zero, i.e., $q_M + q_P = 1$. That is, a user associate only with the strongest BS in terms of received power.

\section{Analysis of Outage Probability and Average Rate}
\label{sec:out_rate}

In this section, we characterize the ${\rm SINR}$ for downlink transmission to a typical user in different modes of operation. Then, we derive closed integral-forms for the outage probability and ergodic rate of downlink transmission for the proposed LA-CTC scheme.

\subsection{${\rm SINR}$ Analysis}

Based on the mode selection  criteria in (\ref{eq:coop_set}), the received signal power at a typical user can be written as
\begin{align}
\underbrace{\sum_{x_i \in \mathcal{B}} \frac{\sqrt{P_i} h_{i,0}}{ \|x_i\|^{\frac{\alpha_i}{2}}} X}_{\text{useful signal}} +  \underbrace{\sum_{j=1}^2  \sum_{x_i \in \Phi_j \setminus \mathcal{B}} \frac{\sqrt{P_j} g_{j,i}}{ \|x_i\|^{\frac{\alpha_j}{2}}} Y_{j,i}}_{\text{inter-cell interference}} + Z  \label{eq:rec_sig}
\end{align}
where $h_{i,0}$ and $g_{j,i}$ are the small-scale fading coefficients for the links between the typical user and the serving and interfering BSs, respectively. $\{ h_{i,0}, g_{j,i} \} \sim \mathcal{CN}(0,1)$ are i.i.d. circular complex Gaussian random variables.  {That is, $|h_{i,0}|$ and $|g_{j,i}|$ are Rayleigh-distributed random variables, the channel power envelope is exponentially-distributed as $\Exp(1)$, and the phase shift in uniformly distributed in $[0,2\pi]$.} $X$ and $Y_{j,i}$ are two zero-mean and unity-variance random variables that represent the jointly transmitted data by  set $\mathcal{B}$ of the serving BSs and the data sent by the interfering BSs, respectively. $Z \sim \mathcal{CN}(0,\sigma_z^2)$ is the additive white noise at the receiver.  {No CSI is assumed at BSs} and that the channel coherence time is greater than or equal to the frame duration. 

 {Note that, ideally, the interference signals received at a user are dependent since interferers could be cooperating as well. However, as can be seen in (\ref{eq:rec_sig}),  $Y_{i,j}$s are independent. The rationale behind this assumption is as follows. Given that two BSs  (at distance $z_1$ and $z_2$ from a location $y$) are cooperating and interfering to a certain user located at $y$, we know that: (a) the received interference power from two cooperating BSs is $|\sqrt{P_1} g_{1} z_1^{-0.5\alpha_1} + \sqrt{P_2} g_{2} z_2^{-0.5\alpha_2}|^2$ which has a Laplace transform (LT)\footnote{The received power from any two cooperating BSs is exponentially-distributed. More details about the distribution are given in \textbf{Appendix \ref{thm1proof}}.} of $\mathcal{L}_{\text{actual}}(s) = (1+(\theta_1^2 + \theta_2^2)s)^{-1}$ where $\theta_i = \sqrt{P_i} z_i^{-0.5\alpha_i}$ , (b) the received interference power is assumed to be $|\sqrt{P_1} g_{1} z_1^{-0.5\alpha_1}|^2 + |\sqrt{P_2} g_{2} z_2^{-0.5\alpha_2}|^2$ that has a LT\footnote{After the assumption, the received power from any two cooperating BSs becomes a sum of two independent exponentially-distributed random variables with different means which is equivalent to a hyperexponential random variable with mean $\theta_1 + \theta_2$ and variance $\theta_1^2 + \theta_2^2$.} of $\mathcal{L}_{\text{assump}}(s) = (1+\theta_2^2s)^{-1} (1+\theta_1^2s)^{-1}$, and (c) the outage probability is a decreasing function of the Laplace transform of the interference (see \textbf{Appendix \ref{thm1proof}}). Hence, $\mathcal{L}_{\text{actual}}(s) \geq \mathcal{L}_{\text{assump}}(s)$ and the independence assumption gives a lower bound on the outage probability.}

Thus, the received ${\rm SINR}$ at a typical receiver is given by
\begin{align}
\rm{SINR}(\mathcal{B}) &= \frac{|\sum_{x_i \in \mathcal{B}} \sqrt{P_i} h_{i,0} \|x_i\|^{-\frac{\alpha_i}{2}}|^2}{\sum_{j=1}^2 P_j \sum_{x_i \in \Phi_j \setminus \mathcal{B}} |g_{j,i}|^2 \|x_i\|^{-\alpha_j} + \sigma_z^2}. \label{eq:SINR}
\end{align}
Note that in (\ref{eq:SINR}), the effect of the user's mode of operation is reflected in $\mathcal{B}$. That is, different modes of operation lead to different levels of the useful signal power (higher/lower) and aggregate interference power (lower/higher). 

 {To show the important role that the proposed mode of operation plays in improving the level of the received ${\rm SINR}$ at the typical user, we consider the following scenario. Considering the no-cooperation case (i.e., when $\beta = 1$) and the cell association based on the strongest signal power regardless of the BS tier,} the serving BS $x$ is selected as follows:
\begin{align}
x &= \arg \max_{x \in \Phi_1 \cup \Phi_2} \{ P_i \|x\|^{-\alpha_i} \}. \label{eq:trad}
\end{align}
In this case, macro users, who are close to the boundaries of the deployed pico cells' coverage, experience high interference, and consequently low ${\rm SINR}$. In the proposed scheme, these users are likely to change their mode of operation to use CoMP transmission instead of single cell transmission. The proposed scheme increases the power level of the useful signal and reduces the total interference power by forcing the strongest interferer to cooperate with the original transmitter. The reduction in the interference power level along with the increase of the useful signal power enhances the received ${\rm SINR}$ at the CoMP user.

\subsection{Outage Probability}

Using the instantaneous ${\rm SINR}$ given in (\ref{eq:SINR}), we can obtain the outage probability $\mathcal{O}$ of the overall system. Here, outage probability is defined as the probability that the received ${\rm SINR}$ is less than a predefined threshold $\tau$. Note that $\tau$ is a design parameter and it is chosen to satisfy certain quality-of-service requirements of users. We denote by $\mathcal{O}_M$, $\mathcal{O}_P$, and $\mathcal{O}_C$  the outage probability of a randomly located user conditioned on its mode of operations, i.e., non-CoMP macro user, non-CoMP pico user, and CoMP user, respectively. For example, the outage probability of a randomly located user given that it operates in the non-CoMP macro mode is obtained by
\begin{align}
\mathcal{O}_M &= \mathbb{E}_x \left[ \mathbb{P} \left[ \rm{SINR}(\mathcal{B} = \{x_1\}) \le \tau \right] \right]. \label{eq:outage_gen}
\end{align}

Since the three modes, i.e., non-CoMP macro mode, non-CoMP pico mode, and CoMP mode, are mutually exclusive, the overall outage probability in the network can be obtained by using the law of total probability as follows:
\begin{align}
\mathcal{O} &= q_M \mathcal{O}_M + q_P \mathcal{O}_P + q_C \mathcal{O}_C \label{eq:tot_out}
\end{align}
where $q_M$, $q_P$, and $q_C$ are given in Lemma \ref{lem1}. The following theorem gives the outage probabilities for a typical user under different modes of operation.

\begin{thm}\normalfont 
The outage probabilities for a typical user given that this user operates as a non-CoMP macro user, or as a non-CoMP pico user, or as a CoMP user are
\begin{align}
&\mathcal{O}_M = 1 - \!\int \limits_{\mathbb{R}_+}\! \exp \Scale[0.91]{\left[\frac{- \tau \sigma_z^2}{P_1 r^{-\alpha_1}} \right] \prod \limits_{j=1}^2 \mathcal{L}_{I_j}\left( \frac{\tau r^{\alpha_1}}{P_1 } \right) f_{R_1}(r)} \text{d}r \label{eq:O_M}\\ 
&\mathcal{O}_P = 1 - \!\int \limits_{\mathbb{R}_+}\! \exp \Scale[0.91]{\left[\frac{- \tau \sigma_z^2}{P_2 r^{-\alpha_2}} \right] \prod \limits_{j=1}^2 \mathcal{L}_{I_j}\left( \frac{\tau r^{\alpha_2}}{P_2 } \right) f_{R_2}(r)} \text{d}r \label{eq:O_P}\\
&\mathcal{O}_C = \nonumber\\
&~~1 - \!\int \limits_{\mathcal{A}}\! \Scale[0.91]{\exp \left[\frac{- \tau \sigma_z^2}{\sum \limits_{i=1}^2 P_i r_i^{-\alpha_i}} \right] \prod \limits_{j=1}^2 \mathcal{L}^\star_{I_j}\left( \frac{\tau}{\sum \limits_{i=1}^2 P_i r_i^{-\alpha_i}} \right) f_{R_C}(\mathbf{r})} \text{d} \mathbf{r} \label{eq:O_C}
\end{align}
where $\mathcal{A}$ is defined in (\ref{eq:A}), $f_{R_1}(r)$, $f_{R_2}(r)$ and $f_{R_C}(\mathbf{r})$ are given in Lemma \ref{lem1}, and
\begin{align}
\Scale[1]{\mathcal{L}_{I_j}\left( \frac{\tau r^{\alpha_i}}{P_i } \right)} &= \Scale[1]{\exp \left[ -2 \pi \lambda_j \left(\tau \frac{P_j}{P_i}\right)^{\frac{2}{\alpha_j}} r^{\frac{2\alpha_i}{\alpha_j}} \mathcal{F}\left((\frac{a_{ij}}{\tau})^{\frac{1}{\alpha_j}} ,\alpha_j\right) \right]} \nonumber \\
\Scale[1]{\mathcal{L}^\star_{I_j}\left( s \right)} &= \Scale[1]{\exp \left[ - 2 \pi \lambda_j (s P_j)^{\frac{2}{\alpha_j}} \mathcal{F}\left((\frac{1}{s P_j})^{\frac{1}{\alpha_j}} r_j,\alpha_j\right)\right]} \nonumber\\
a_{ij} &= \left\{
\begin{array}{l l}
\beta, & i = 1 \text{~and~} j=2 \\
1, & \text{otherwise}
\end{array}\right. \nonumber \\
\mathcal{F}(y,\alpha) &= \int_{y}^\infty \frac{u}{1+u^{\alpha}} \text{d}u. \label{eq:F}
\end{align}
\label{thm1}
\begin{IEEEproof}
See \textbf{Appendix \ref{thm1proof}}.
\end{IEEEproof}
\end{thm}

\textbf{Theorem \ref{thm1}} provides general closed integral-form expressions for the outage probabilities for a randomly located user. Note that the function $\mathcal{F}(y,\alpha)$ can be evaluated numerically. Furthermore, in some special cases $\mathcal{F}(y,\alpha)$ reduces to simple closed-form expressions (see \textbf{Appendix \ref{F_fun}}). The expressions in \textbf{Theorem \ref{thm1}} can be used to obtain the performances for some special cases by varying $\beta$, $\alpha_i$, and cross-interference mitigation scheme. 

 {In the following, we introduce three main schemes, namely, range expansion (RE), full cooperation (FC), and two-tier network with strongest BS association and no cooperation (Tr) schemes, which will be compared to our proposed scheme. }
\vspace{0.3cm}
\subsubsection{Non-cooperative two-tier cellular network with range expansion (RE)}
\label{sec:re}
 {RE is a non-cooperative scheme in which the association to the pico-tier  is biased such that some macro users are offloaded to the strongest pico BS even though the received power from this pico BS is less than that from the macro BS, hence, the range of the picocell is expanded. To elaborate, in Fig. \ref{map}, the cooperation regions of our proposed scheme become a part of the pico BSs' coverage areas and users in these regions become pico users. In other words, it can be seen that RE offloads each CoMP user to its strongest pico BS where these users switch to the non-CoMP mode. That is, the positive bias to the pico-tier association becomes $\beta$. Note that, $\beta$ refers to both cooperation threshold of LA-CTC scheme and bias factor of RE scheme depending on the context.} For the described scheme, according to \textbf{Theorem \ref{thm1}}, the outage of the macro-tier (as given in (\ref{eq:O_M})) remains unchanged, where the outage of the pico-tier can be obtained as in the following corollary.
\begin{col}\normalfont 
({\it Range expansion}) In the special case of a non-cooperative two-tier network with a biased association to the pico-tier, the outage probability of a randomly located pico user is given by 
\begin{align}
\mathcal{O}_P^{RE} &= 1 - \!\int \limits_{\mathbb{R}_+}\! \exp \Scale[0.9]{\left[\frac{- \tau \sigma_z^2}{P_2 r^{-\alpha_2}} \right] \prod \limits_{j=1}^2 \mathcal{L}_{I_j}\left( \frac{\tau  r^{\alpha_2}}{P_2} \right) f_{R_2}^{RE}(r)} \text{d}r
\end{align}
where
\begin{align}
f_{R_2}^{RE}(x) = \ \frac{2\pi \lambda_2}{q_P^{ER}} r \exp \Resize{4.65cm}{\left[ -\pi \left( \lambda_2 r^2 + \lambda_1 \left(\frac{P_1}{\beta P_2}\right)^{\frac{2}{\alpha_1}} r^{\frac{2\alpha_2}{\alpha_1}} \right) \right]} \nonumber
\end{align}
in which ${\mathcal{L}_{I_j}( \cdot )}$ is given in \textbf{Theorem \ref{thm1}} with $a_{ij} = \beta^{j-i}$ and $q_P^{ER} = q_P + q_C$.
\label{col1}
\begin{IEEEproof}
We follow the same proofs as in \textbf{Appendix \ref{thm1proof}} and  \textbf{Appendix \ref{lem1proof}}, respectively, while replacing $P_2$ by $\beta P_2$. 
\end{IEEEproof}
\end{col}

 {Hence, the overall outage probability of RE is given by
\begin{align}
\mathcal{O}^{RE} &= q_M \mathcal{O}_M + q_P^{RE} \mathcal{O}_P^{RE} \label{eq:out_re}
\end{align}
where $\mathcal{O}_M$ is given by (\ref{eq:O_M}), and $q_P^{RE} = q_P + q_C$ where $q_P$ and $q_C$ are given in \textbf{Lemma \ref{lem1}}.}

In this case, the closest interferer from the macro-tier to a typical pico user is at least at a distance of $\left(\frac{P_1}{\beta P_2}\right)^{\frac{1}{\alpha_1}} r_2^{\frac{\alpha_2}{\alpha_1}}$ instead of $  r_2$. That is, the macro BS corresponding to which the received power at the pico user is the highest is considered as the closest interferer. Furthermore, for the pico user in the expanded picocell coverage area, cf. Fig. \ref{map}, the highest interference signal from the macro-tier is even higher than its useful signal received from the serving pico BS. This means that the {\rm SINR} of this user is less than $1$. This implies that the RE scheme degrades system performance compared to our proposed scheme. 

\vspace{0.3cm}
\subsubsection{Fully-cooperative two-tier cellular network (FC)}
\label{sec:fc}
In this scheme, any typical user, regardless of its location, connects to the strongest BS from each tier, i.e., all users operate in the CoMP mode. The outage probability in this case is provided in the following corollary.
\begin{col}\normalfont 
({\it Full cooperation}) In the special case of a fully-cooperative two-tier cellular network, the overall outage probability of the network is given by
\begin{align}
&\mathcal{O}^{FC}  \!=  \nonumber\\
&\!1 - \!\int \limits_{\mathbb{R}_+^2}\! \Scale[0.9]{\exp \left[\frac{- \tau \sigma_z^2}{\sum \limits_{i=1}^2 P_i r_i^{-\alpha_i}} \right] \prod \limits_{j=1}^2 \mathcal{L}^\star_{I_j}\left( \frac{\tau}{\sum \limits_{i=1}^2 P_i r_i^{-\alpha_i}} \right) \prod \limits_{j=1}^2 f_{R_j}'(r_j)} \text{d} \mathbf{r} \label{eq:O_full}
\end{align}
where $f_{R_j}'(r_j) $ is given in (\ref{eq:distance_dist}) and $\mathcal{L}^\star_{I_j}(\cdot) $ is given in \textbf{Theorem \ref{thm1}}.
\label{col2}
\begin{IEEEproof}
We use the fact that the BS with the strongest received signal at the typical user from the $i^{th}$ tier is the nearest BS to this typical user among all BSs in this tier. That is, the distance to the strongest BS is Rayleigh distributed, i.e., $f_{R_j}'(r_j) = 2 \pi \lambda_j r_j \exp[{-\pi \lambda_j r_j^2}]$ and the joint PDF of the distance is the multiplication of the two distributions because of the independence between the two random variables. By plugging the PDF of the distance in (\ref{eq:O_C_def}) and following the proof of $\mathcal{O}_C$ in \textbf{Appendix \ref{thm1proof}}, we obtain the results in (\ref{eq:O_full}) where $q_C = 1$ and $q_M = q_P = 0$.
\end{IEEEproof}
\end{col}

 {In this case, the closet interferers from the macro-tier and the pico-tier to any user is at least at a distance of $r_1$ and $r_2$, respectively. Hence, the performance of all users is improved and the overall outage performance is better compared to the LA-CTC scheme, however, this enhancement comes at the expense of the overhead due to data exchange between the two cooperating BSs.}

\vspace{0.3cm}
\subsubsection{Interference-limited traditional two-tier cellular network (Tr)}
\label{sec:tr}
In this case, each user associates with the strongest BS from any tier as defined in (\ref{eq:trad}). The outage probability can be obtained from \textbf{Theorem \ref{thm1}} as in the following corollary.
\begin{col}\normalfont 
({\it No cooperation with strongest BS association}) In the special case of a two-tier cellular network when each user associates with the BS that results in the highest average received power, the total outage probability simplifies to
\begin{align}
\mathcal{O}^{Tr} &= 1 - \frac{1}{ 1 + 2 \tau^{\frac{2}{\alpha}}  \mathcal{F}\left(\tau^{-\frac{1}{\alpha}},\alpha \right)} \label{eq:O_M_no_noise}
\end{align}
where the network operates in the interference-limited regime and $\alpha_1 = \alpha_2 = \alpha$.
\label{col3}
\begin{IEEEproof}
By using the results in \textbf{Theorem \ref{thm1}} and substituting $\alpha_1 = \alpha_2 = \alpha$, $\beta = 1$ and $\sigma_z^2 = 0$, we obtain $\mathcal{O}_M^{Tr} = \mathcal{O}_P^{Tr}$ (\ref{eq:O_M}), $q_C = 0$, and $q_M$ and $q_P$ are as given in (\ref{eq:q_same}). Then, the overall outage probability is obtained as in (\ref{eq:O_M_no_noise}).
\end{IEEEproof}
\end{col}

 {In this scheme, the closest interferer from the pico-tier to a typical macro user is at least at a distance $R_1$ compared to $ \left(\frac{\beta P_2}{P_1}\right)^{\frac{1}{\alpha_2}} R_1^{\frac{\alpha_1}{\alpha_2}}$ in the case of LA-CTC scheme. That is, the strongest interferer is closer to the user which degrades the overall performance compared to our proposed scheme.} It can be seen that, in this case, the outage probability is independent of the BS intensity and transmit power. That is because, the association is based on the highest signal received from any BS which means that the outage probability does not change when more BSs are deployed or  the transmit power is increased while assuming   the same path-loss exponent. 

 {{Note that the results presented in \textbf{Corollaries \ref{col1}}, \textbf{\ref{col2}}, and \textbf{\ref{col3}} are consistent with the previous results in \cite{6287527, martin2013comp, dhillon2012modeling} on multi-tier cellular networks. Furthermore, the same result in \textbf{Corollary \ref{col3}} can be obtained for the non-cooperative single-tier cellular case by substituting $\lambda = \lambda_1 + \lambda_2$ and assuming that both tiers are identical in powers ($P_i = P$) and path-loss exponents ($\alpha_i = \alpha$), or simply substituting $\lambda_2 = 0$. This result is consistent with the previous results on single-tier networks in \cite{165165}.}}

\subsection{Average Ergodic Rate}
\label{sec:rate}

Based on the  {conditional} outage probabilities defined in \textbf{Theorem \ref{thm1}},  {we derive  expressions of the ergodic rates for a typical user when it operates in different modes}. The ergodic rate is measured in nats/sec/Hz where it represents the spectral efficiency of transmission to a user. Using the independence property used in (\ref{eq:tot_out}), the average ergodic rate for a user is given by
\begin{align}
\mathcal{R} &= q_M \mathcal{R}_M + q_P \mathcal{R}_P + q_C \mathcal{R}_C \label{eq:tot_rate}
\end{align}
where $\mathcal{R}_M$, $\mathcal{R}_P$, and $\mathcal{R}_C$ are the ergodic rate of a typical user given that it operates in the non-CoMP mode and served by a macro BS,  in the non-CoMP mode and served by a pico BS, and in the CoMP mode, respectively, and the association probabilities are given in \textbf{Lemma \ref{lem1}}. 

In the following theorem, we derive an expression for the ergodic rate of a randomly located CoMP user. Note that the rate of non-CoMP users, i.e., macro or pico users, can be obtained following the same procedure and the overall average ergodic rate for a user in the network can be obtained from (\ref{eq:tot_rate}). The expressions for the ergodic rate of downlink transmission for the RE, FC, and Tr schemes follow the same procedure.

\begin{thm}\normalfont 
The ergodic rate for a typical CoMP user is 
\begin{align}
\mathcal{R}_C &= \!\int \limits_{\mathbb{R}_+}\! 1 - \left[ \mathcal{O}_C\right]_{\tau = e^t - 1} \text{d}t \label{eq:rate}
\end{align}
where $\mathcal{O}_C$ is given in (\ref{eq:O_C}) and $\left[ \cdot \right]_{\tau = f(t)}$ means replacing each $\tau$ by $f(t)$ for some function $f: t \rightarrow \tau$.
\label{thm2}
\begin{IEEEproof}
The ergodic rate for a randomly located CoMP user is defined as
\begin{align}
\mathcal{R}_C &= \mathbb{E}_{\mathbf{r}} \left[ \mathbb{E}_{\rm{SINR}} \left[ \ln (1+\rm{SINR}(\mathcal{B} = \{x_1,x_2\})) \right] \right]\nonumber
\end{align}
where the expectation is taken with respect to the distance between the user and its serving BSs. That is, the ergodic rate can be rewritten as
\begin{align}
\mathcal{R}_C	&= \int \limits_{\mathcal{A}} \mathbb{E}_{\rm{SINR}} \left[ \ln (1+\rm{SINR}(\mathcal{B})) \right]   f_{R_C}(\mathbf{r}) \text{d}\mathbf{x} \nonumber\\
	&= \int \limits_{\mathbb{R}_+} \bigg[\int \limits_{\mathcal{A}} \mathbb{P} \left[\ln (1+\rm{SINR}(\mathcal{B}))  > t \right]   f_{R_C}(\mathbf{r})  \text{d}\mathbf{r}\bigg] ~\text{d}t\nonumber\\
	&= \int \limits_{\mathbb{R}_+} \bigg[\int \limits_{\mathcal{A}} \mathbb{P} \left[\rm{SINR}(\mathcal{B}) > e^t - 1 \right]   f_{R_C}(\mathbf{r})  \text{d}\mathbf{r}\bigg] ~\text{d}t\nonumber
\end{align}
and by using the the definition of $\mathcal{O}_C$ given in (\ref{eq:O_C_def}), we obtain the result in (\ref{eq:rate}).
\end{IEEEproof}
\end{thm}

\section{Numerical Results and Discussion}
\label{sec:results}

\subsection{Performance Metrics and Values of System Parameters}

In this section, we compare the proposed LA-CTC scheme with three schemes in the literature discussed in Sections \ref{sec:re}, \ref{sec:fc}, and \ref{sec:tr}. The first scheme is the flexible cell association, which is referred to as Range Expansion (RE), where its overall outage probability is defined in (\ref{eq:out_re}). In the second scheme, referred to as Full Cooperation (FC), each user in the network is served by two BSs. That is, each user is connected to one BS from each tier that results in the strongest average received power. The overall outage probability for this scheme $\mathcal{O}^{FC}$ is given in \textbf{Corollary \ref{col2}}. Finally, the third scheme is the traditional scheme (Tr) for a two-tier cellular network in which a typical user is served only by the strongest BS and no biasing is used (i.e., $\beta = 0$ dB). The overall outage probability for this scheme $\mathcal{O}^{Tr}$ is given in \textbf{Corollary \ref{col3}}.

The comparison is performed in terms of outage probability, spectral efficiency, and load per BS. {While the first two metrics have been defined before, the load per BS is defined as the average number of users connected to a BS in any tier. Using the independence assumption between point processes of BSs and users, the load per BS for the four schemes can be obtained as given in Table \ref{load_table}.}

{\renewcommand{\arraystretch}{1.8}
\begin{table}[!ht]\small
\centering
\caption{Load per BS for the considered schemes: LA-CTC, Range Expansion, Full Cooperation, and Traditional}
\label{load_table}
\begin{tabular}{| l | l | l| }
\hline
\textbf{Scheme} & \textbf{Macro BS} & \textbf{Pico BS}\\
\hline
\hline
LA-CTC & $\frac{\lambda_u}{\lambda_1} (q_M + q_C)$ & $\frac{\lambda_u}{\lambda_2} (q_P + q_C)$\\
\hline
RE & $\frac{\lambda_u}{\lambda_1} q_M$ & $\frac{\lambda_u}{\lambda_2} (q_P + q_C)$\\
\hline
FC & $\frac{\lambda_u}{\lambda_1} $ & $\frac{\lambda_u}{\lambda_2} $\\
\hline
Tr & $\frac{\lambda_u}{\lambda_1} (q_M + q_C)$ & $\frac{\lambda_u}{\lambda_2} q_P$\\
\hline
\end{tabular}
\end{table}
}
    
For the numerical evaluation, the transmit powers of a macro BS and a pico BS are assumed to be $37$ dBm and $20$ dBm, respectively, while the thermal noise power $\sigma_z^2$ is $-104$ dBm. Independent and identically distributed (i.i.d.) circular complex random variables with zero mean and unit variance are considered to simulate the channels. The macro-tier has an intensity of $\lambda_1 = (500^2 \pi)^{-1}$. Unless otherwise stated, the intensity of BSs in the pico-tier is $5$ times that of the macro-tier, i.e., $\lambda_2 = 5 (500^2 \pi)^{-1}$ and the intensity of users $\lambda_u = 10 (500^2 \pi)^{-1}$. For the evaluation of outage probability, the threshold $\tau$ is set to $0$ dB.

\subsection{Validation of Analysis}
\begin{figure}[!t]
\centering
\includegraphics[width=0.45\textwidth]{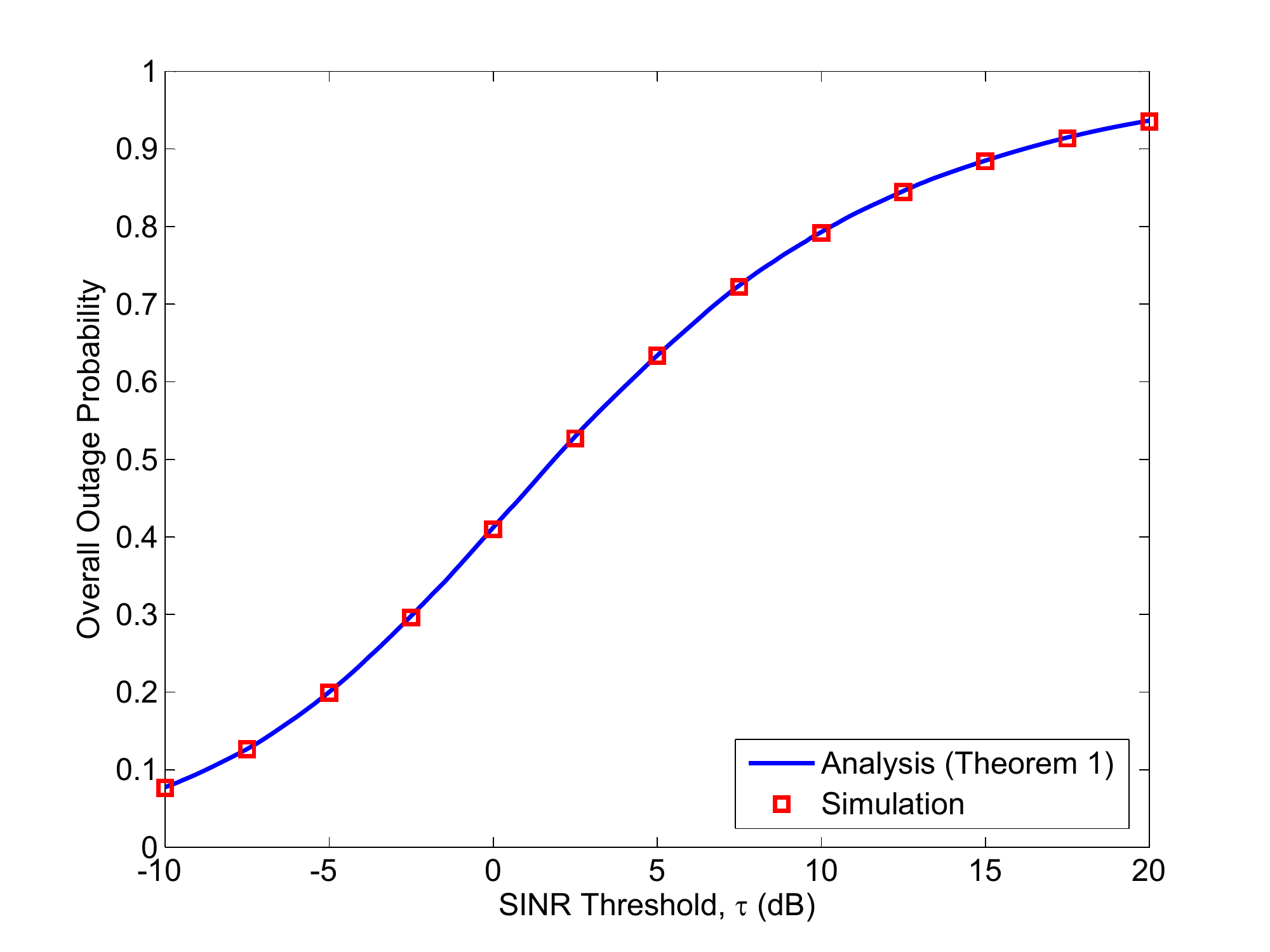}
\caption{Analysis vs. simulation: Overall outage probability for the LA-CTC scheme where $\lambda_1 = (500^2 \pi)^{-1}$, $\lambda_2 = 5 (500^2 \pi)^{-1}$, $P_1 = 37$ dBm, $P_2 = 20$ dBm, $\beta = 4$ dB, $\alpha_1 = \alpha_2 = 4$, and $\sigma_z^2 = -104$ dBm.}
\label{sim_vs_ana}
\end{figure}

In Fig. \ref{sim_vs_ana}, we validate our analysis by comparing the overall outage probability (i.e., CCDF of ${\rm SINR}$ at $\tau$) for the  LA-CTC scheme obtained from both the analysis (\ref{eq:tot_out}) and simulation.  {Monte Carlo simulations via MATLAB are used where the simulation area is $10$km $\times 10$km and the results are averaged over $10^6$ iterations. In each realization, the performance is evaluated for a typical user at the origin where the BSs are deployed according to two independent PPPs.} It can be seen that the analytical results (see the expressions given in (\ref{eq:tot_out}) and \textbf{Theorem \ref{thm1}}) match exactly with the simulation results for all ${\rm SINR}$ thresholds which reflects the accuracy of our analysis. Therefore, from now and on, we use the analytical expressions to evaluate the system performance. 

\subsection{Outage Probability}
\begin{figure}[!t]
\centering
\includegraphics[width=0.45\textwidth]{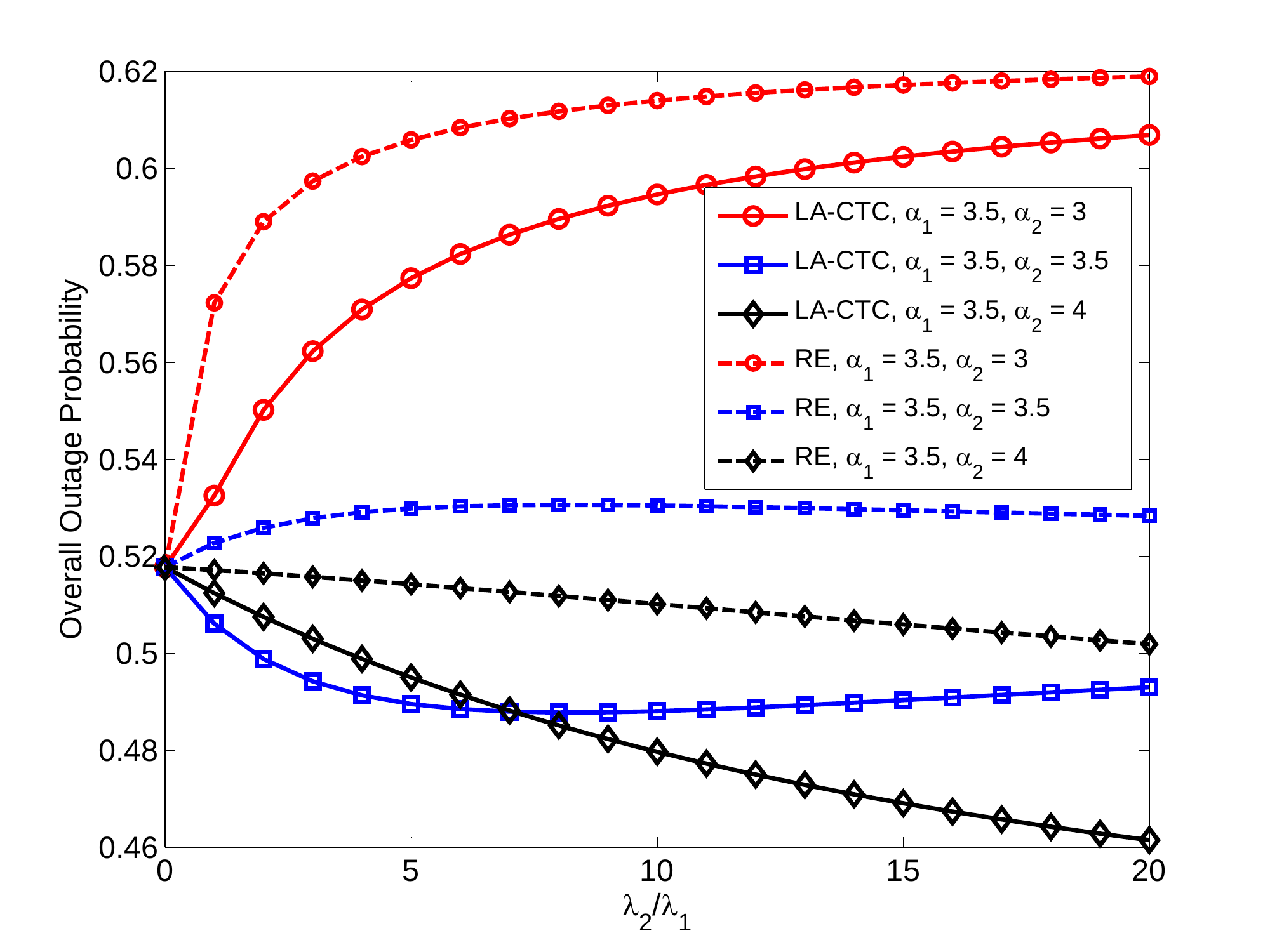}
\caption{LA-CTC vs. Range Expansion: Overall outage probability for different path-loss exponents vs. the ratio of BS intensity where $\lambda_1 = (500^2 \pi)^{-1}$, $P_1 = 37$ dBm, $P_2 = 20$ dBm, $\beta = 4$ dB, and $\sigma_z^2 = -104$ dBm.}
\label{overall_outage_vs_lambda}
\end{figure}

Fig. \ref{overall_outage_vs_lambda} shows the effect of varying both the path-loss exponents and the BS intensity on the overall outage probabilities for the LA-CTC  and RE schemes. From this figure, it can be seen that the proposed LA-CTC scheme has two advantages over the RE scheme. Firstly, the overall outage probability for the LA-CTC scheme is better than that for RE scheme for all the different values of path-loss exponents. Furthermore, in some cases, e.g., when $\alpha_1 = \alpha_2$, with the RE scheme, the outage probability deteriorates with increasing pico BS intensity, while with the LA-CTC scheme the outage probability improves  under the same conditions. The proposed scheme outperforms the RE scheme since for a macro user it eliminates the  highest interferer from the pico-tier when the highest received interference power is within a predefined range, i.e., $\frac{1}{\beta} P_2 R_2^{-\alpha_2} < P_{int} < P_2 R_2^{-\alpha_2}$. Moreover, it uses this interfering BS as a cooperation partner along with the original serving macro BS to serve this user.


For the case when $\alpha_1 = \alpha_2$, while using the RE scheme, increasing the pico BS intensity limits the effect of the thermal noise and the network operates in the interference-limited regime in which the inter-BS interference dominates the performance. Consequently, the outage probability remains constant when the pico BS intensity is high enough to cancel the effect of both the biasing and the thermal noise. On the hand, the outage probability for the LA-CTC scheme is improved for the same case (i.e., when $\alpha_1 = \alpha_2$), because the proposed scheme mitigates the highest interferer from the pico BS and also uses it as a serving transmitter.  For the case when $\alpha_2$ is higher than $\alpha_1$, the pico BSs become more isolated from the the macro BSs which, in turn, reduces the effect of interference and improves the overall outage probability for the LA-CTC and RE schemes. However, the improvement in outage due to the LA-CTC scheme is much higher than that due to the RE scheme because of the same reason mentioned in the previous case. Finally, in the case when $\alpha_2$ is less than $\alpha_1$, the outage performance deteriorates for the two schemes. However, the proposed LA-CTC scheme limits the performance loss by using cooperation between the original serving BS and its highest interferer from the other tier to serve the  user in CoMP mode.

\begin{figure}[!t]
\centering
\includegraphics[width=0.45\textwidth]{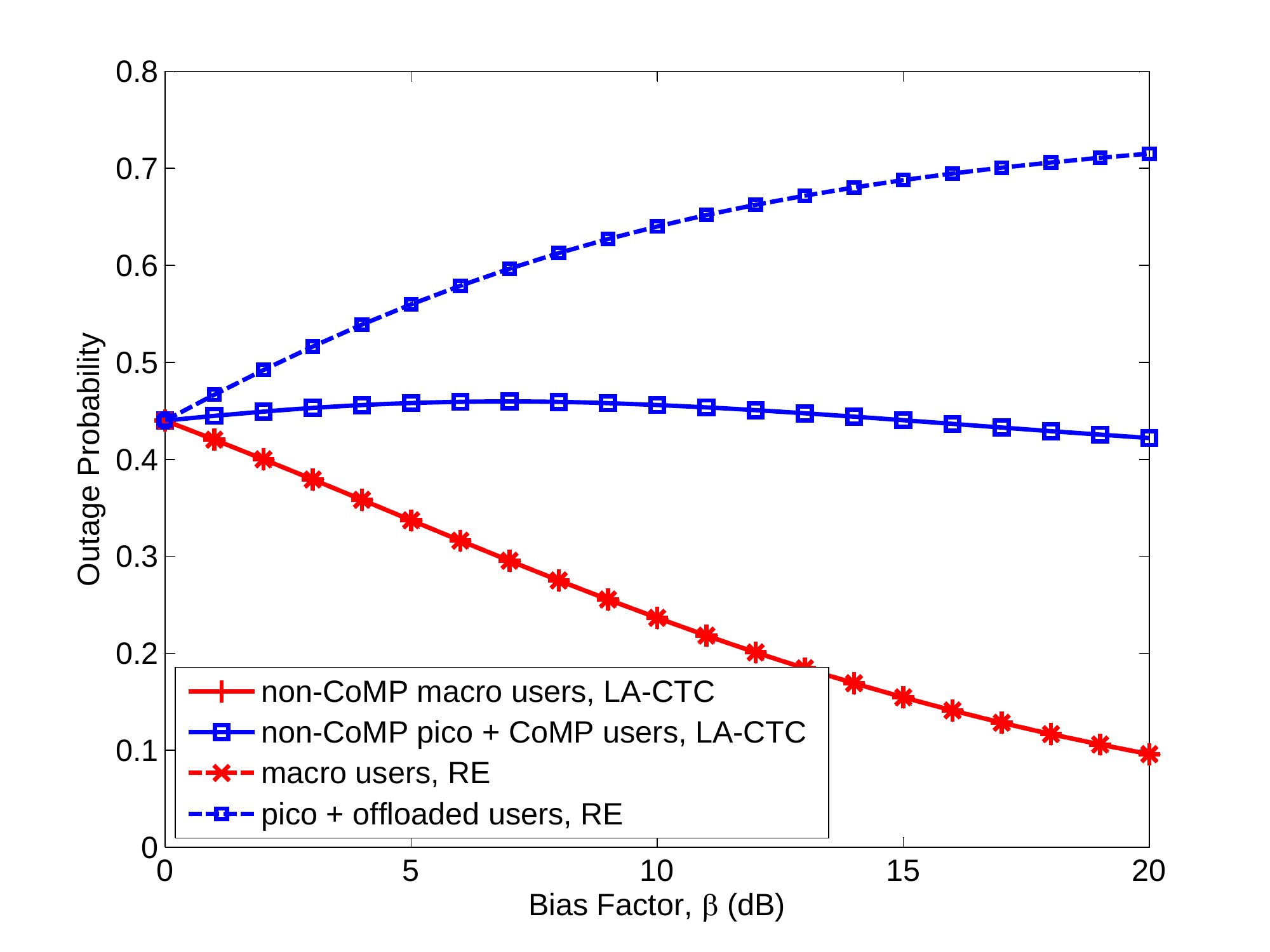}
\caption{LA-CTC vs. Range Expansion: Outage probability vs. the cooperation threshold (bias factor) $\beta$ where $\lambda_1 = (500^2 \pi)^{-1}$, $\lambda_2 = 5 (500^2 \pi)^{-1}$, $P_1 = 37$ dBm, $P_2 = 20$ dBm, $\alpha_1 = \alpha_2 = 4$, and $\sigma_z^2 = -104$ dBm.}
\label{outage_vs_beta}
\end{figure}

 {Fig. \ref{outage_vs_beta} depicts the effect of increasing the cooperation threshold (bias factor) $\beta$ on the outage performance of  each operation mode for the proposed scheme and the RE scheme. Since the outage probability of the offloaded users is added to the outage of the pico users in the RE scheme, for  a fair comparison, in Fig. \ref{outage_vs_beta} we add the outage of CoMP users in the LA-CTC scheme to the pico users' outage as well (i.e., $q_P \mathcal{O}_P + q_C \mathcal{O}_C$).} In Fig. \ref{outage_vs_beta}, from the perspective of macro users, as the cooperation threshold (bias factor) increases, both schemes improve the outage performance compared to the Tr scheme (i.e., when $\beta = 0$ dB). This improvement is due to offloading macro users with poor ${\rm SINR}$ conditions to the pico-tier (in the RE scheme) or to the CoMP transmission mode (in the LA-CTC scheme). Although offloading users improves the outage of the macro-tier in the RE scheme, it degrades the outage of the pico-tier and the overall network as shown in Fig. \ref{outage_vs_beta}. This degradation in outage occurs because each offloaded user connects to a pico BS that does not result in the strongest received power; hence, the user's ${\rm SINR}$ deteriorates. On the other hand, in the LA-CTC scheme, CoMP users are served by both the BSs which boosts the ${\rm SINR}$ of these users and compensates for the loss incurred in the RE scheme. That is, the LA-CTC scheme provides a better outage for the CoMP users compared to the offloaded users in the RE scheme while maintaining the same macro-tier performance.

\begin{figure}[!t]
\centering
\includegraphics[width=0.45\textwidth]{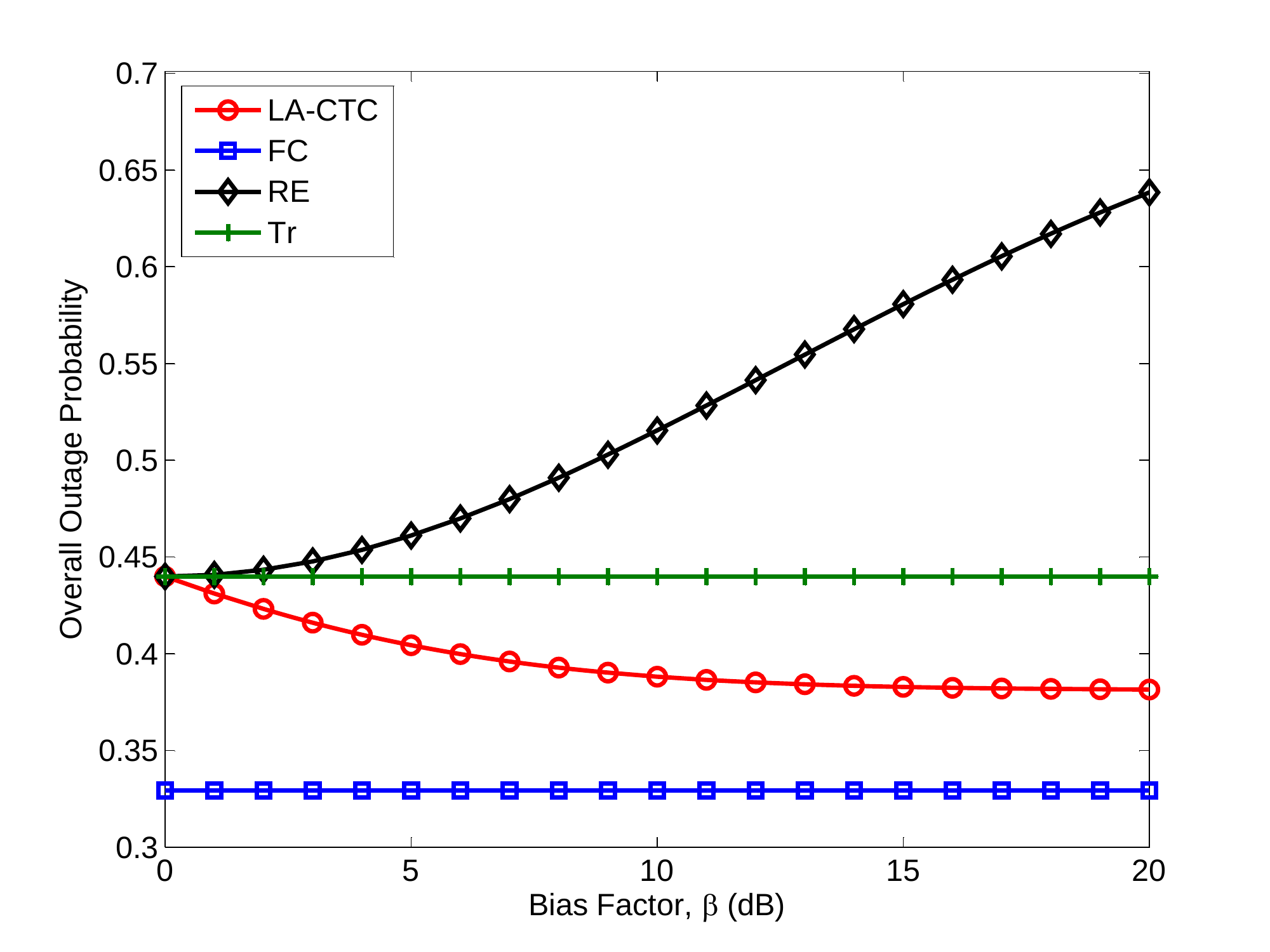}
\caption{LA-CTC vs. Traditional, Range Expansion, and Full Cooperation: Overall outage probability vs. the cooperation threshold (bias factor) $\beta$ where $\lambda_1 = (500^2 \pi)^{-1}$, $\lambda_2 = 5 (500^2 \pi)^{-1}$, $P_1 = 37$ dBm, $P_2 = 20$ dBm, $\alpha_1 = \alpha_2 = 4$, and $\sigma_z^2 = -104$ dBm.}
\label{overall_outage_vs_beta}
\end{figure}

In Fig. \ref{overall_outage_vs_beta}, it can be seen that the overall outage probability of the proposed scheme lies between those of the traditional and the full cooperation schemes. Furthermore, compared to the RE scheme, the proposed scheme significantly improves the overall outage probability of the system. As the cooperation threshold (bias factor) increases, more users are served via cooperation and the performance of the LA-CTC scheme approaches that of the FC scheme. When $\beta \rightarrow \infty$, the gap between the two curves results from the outage of the non-CoMP pico users which is not affected by increasing the bias factor. On the other hand, the gap between the  {performance} of the RE scheme and the Tr scheme increases when the bias factor increases. This is because, a higher $\beta$ causes more users to be offloaded to the pico-tier and served with ${\rm SINR}$ less than $0$ dB, hence, the overall outage probability deteriorates. That is, the LA-CTC scheme outperforms both the Tr scheme and the RE scheme in terms of overall outage probability while approaching the performance of the full cooperation scheme.

\subsection{Spectral Efficiency}
\begin{figure}[!t]
\centering
\includegraphics[width=0.45\textwidth]{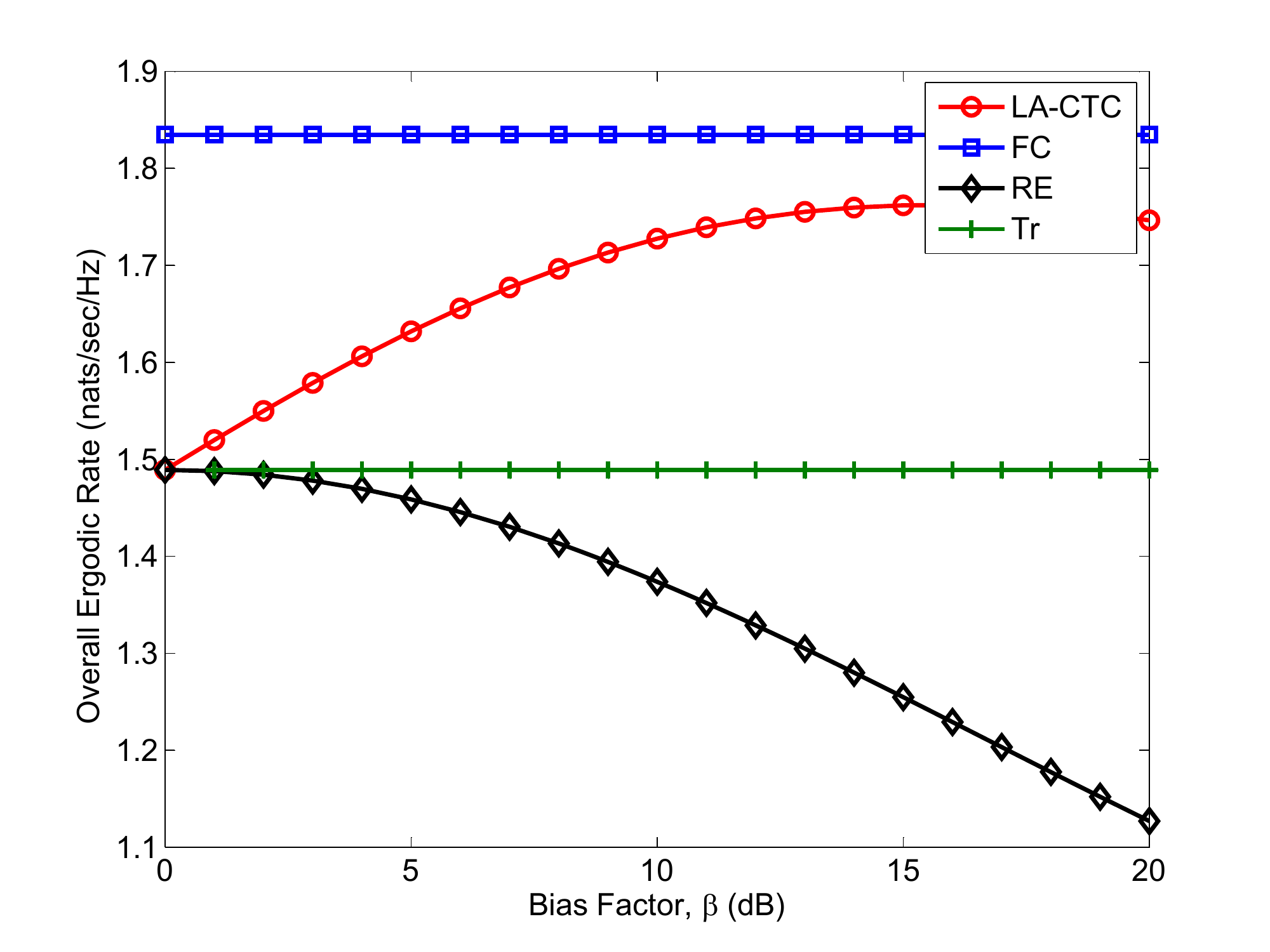}
\caption{LA-CTC vs. Traditional, Range Expansion, and Full Cooperation: Overall average ergodic rate vs. the cooperation threshold (bias factor) $\beta$ where $\lambda_1 = (500^2 \pi)^{-1}$, $\lambda_2 = 5 (500^2 \pi)^{-1}$, $P_1 = 37$ dBm, $P_2 = 20$ dBm, $\alpha_1 = \alpha_2 = 4$, and $\sigma_z^2 = -104$ dBm.}
\label{overall_rate_vs_beta}
\end{figure}

In terms of the overall average achievable rate, it can be seen in Fig. \ref{overall_rate_vs_beta} that the LA-CTC scheme improves the performance of the network compared to the Tr scheme as the cooperation threshold increases. This result is consistent with that in Fig. \ref{outage_vs_beta}. By using CoMP transmission, the proposed scheme increases the ${\rm SINR}$ of users who receive high interference from the pico-tier,  by increasing the useful signal power along with decreasing the interference power. On the other hand, as the bias factor increases, the overall average ergodic rate deteriorates with the RE scheme compared to both the Tr and LA-CTC schemes. This is also consistent with the results in  Fig. \ref{outage_vs_beta} since the offloaded users have lower ${\rm SINR}$ compared to that they had before the offloading. As expected, the full cooperation scheme offers the highest achievable data rate, however, the data rate offered by the LA-CTC scheme approaches that of the full cooperation scheme when the value of the cooperation threshold is high enough.

\begin{figure}[!t]
\centering
\includegraphics[width=0.45\textwidth]{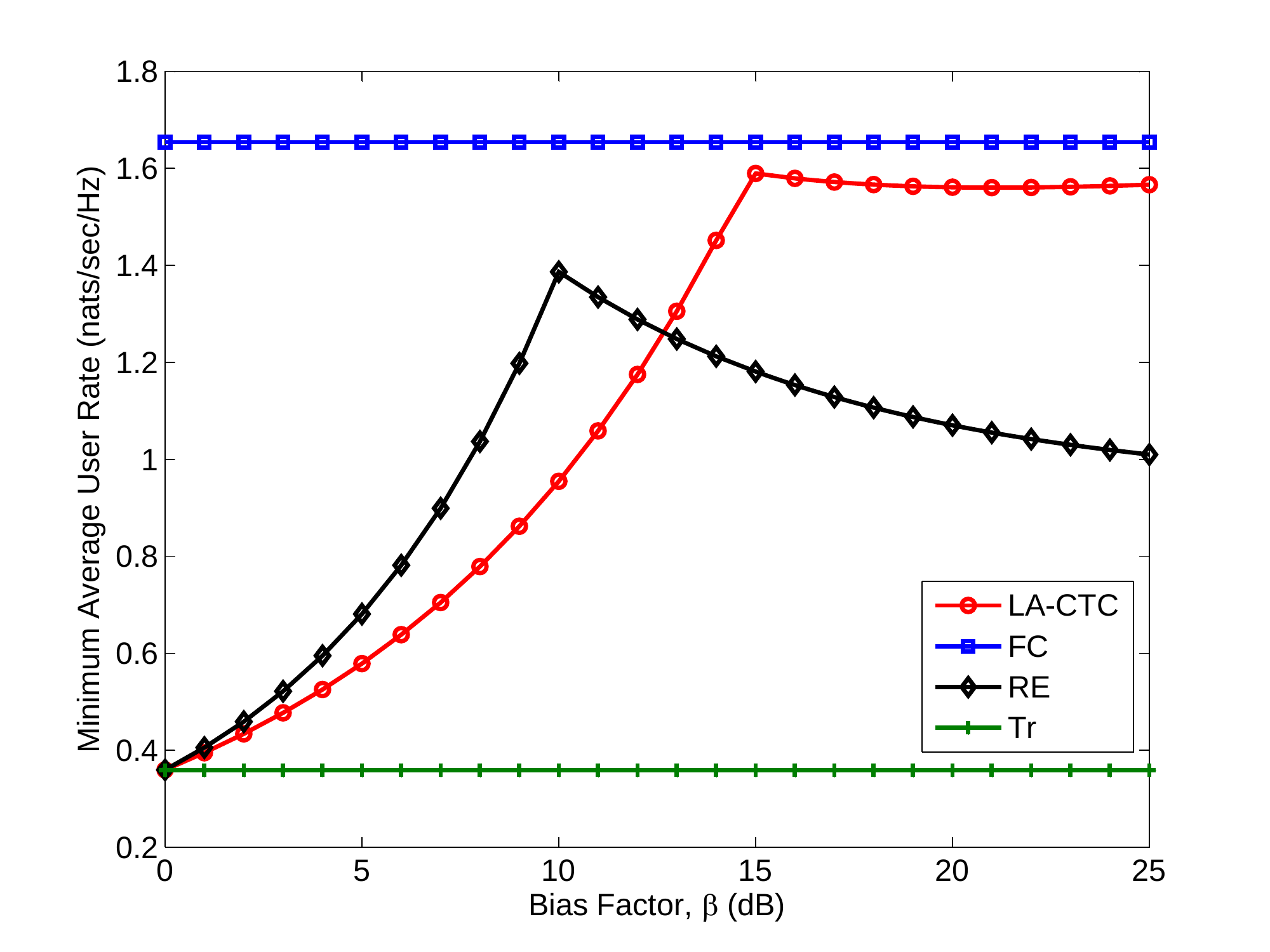}
\caption{LA-CTC vs. Range Expansion: Minimum average user rate for different BS intensities vs. the cooperation threshold (bias factor) $\beta$ where $\lambda_1 = (500^2 \pi)^{-1}$, $\lambda_2 = 5(500^2 \pi)^{-1}$, $\lambda_u = 10 (500^2 \pi)^{-1}$, $P_1 = 37$ dBm, $P_2 = 20$ dBm, $\alpha_1 = \alpha_2 = 4$, and $\sigma_z^2 = -104$ dBm.}
\label{min_avg_rate_vs_beta}
\end{figure}

 {In order to show the impact of using the different schemes on the rate of the legacy users, Fig. \ref{min_avg_rate_vs_beta} compares the performance of the LA-CTC scheme to that of RE scheme in terms of the minimum average ergodic rate the network can provide to a user by any of its tiers. The minimum average user rate offered by a certain BS can be defined as the ratio of the average ergodic rate defined in Section \ref{sec:rate} to the number of users per this BS defined in Table \ref{load_table}. For example, the minimum average rate offered by a macro BS to its users when adopting the LA-CTC scheme is obtained as $\frac{q_M \mathcal{R}_M + q_C \mathcal{R}_C}{q_M + q_C} \frac{\lambda_1}{\lambda_u (q_M + q_C)}$ where the minimum rate offered by a pico BS is $\frac{q_P \mathcal{R}_P + q_C \mathcal{R}_C}{q_P + q_C} \frac{\lambda_2}{\lambda_u (q_P + q_C)}$. Hence, the minimum average rate offered by the network for the LA-CTC scheme can be obtained as
\begin{align}
\min \left\{\frac{q_M \mathcal{R}_M + q_C \mathcal{R}_C}{(q_M + q_C)^2} \frac{\lambda_1}{\lambda_u }, \frac{q_P \mathcal{R}_P + q_C \mathcal{R}_C}{ (q_P + q_C)^2} \frac{\lambda_2}{\lambda_u}\right\}.
\end{align}
Similarly, the minimum average user rate offered by the network for the RE scheme can be obtained by as
\begin{align}
\min \left\{\frac{\mathcal{R}_M}{q_M} \frac{\lambda_1}{\lambda_u},  \frac{\mathcal{R}_P^{RE}}{q_P + q_C} \frac{\lambda_2}{\lambda_u}\right\}.
\end{align}
For the Tr scheme, the minimum rate is equal to that of RE scheme when $\beta$ goes to $0$, while for the FC scheme, it is equal to that of LA-CTC when $\beta$ approaches infinity.}

 {It can be seen in Fig. \ref{min_avg_rate_vs_beta} that, for the RE scheme, as the bias factor increases, the average user rate offered by the network improves up to a maximum point. After this point, the rate offered by pico BSs starts to limit the network performance due to the increase in the number of users per pico BS, hence, the minimum rate starts to degrade. This effect is less severe in the LA-CTC scheme as the increase in the number of users per pico BS due to the increase in the cooperation threshold is compensated by the improvement in the overall rate of the CoMP users offered by the network. That is, the minimum average user rate remains almost constant for high bias factor values. It can also be seen that the performance of the proposed scheme approaches the performance due to full cooperation when $\beta$ is high enough. In addition, the minimum rate offered by each of the LA-CTC  and RE schemes is better than that of the Tr scheme for all $\beta > 0$.}
 
\subsection{Average Load per BS}

\begin{figure}[!t]
\centering
\includegraphics[width=0.45\textwidth]{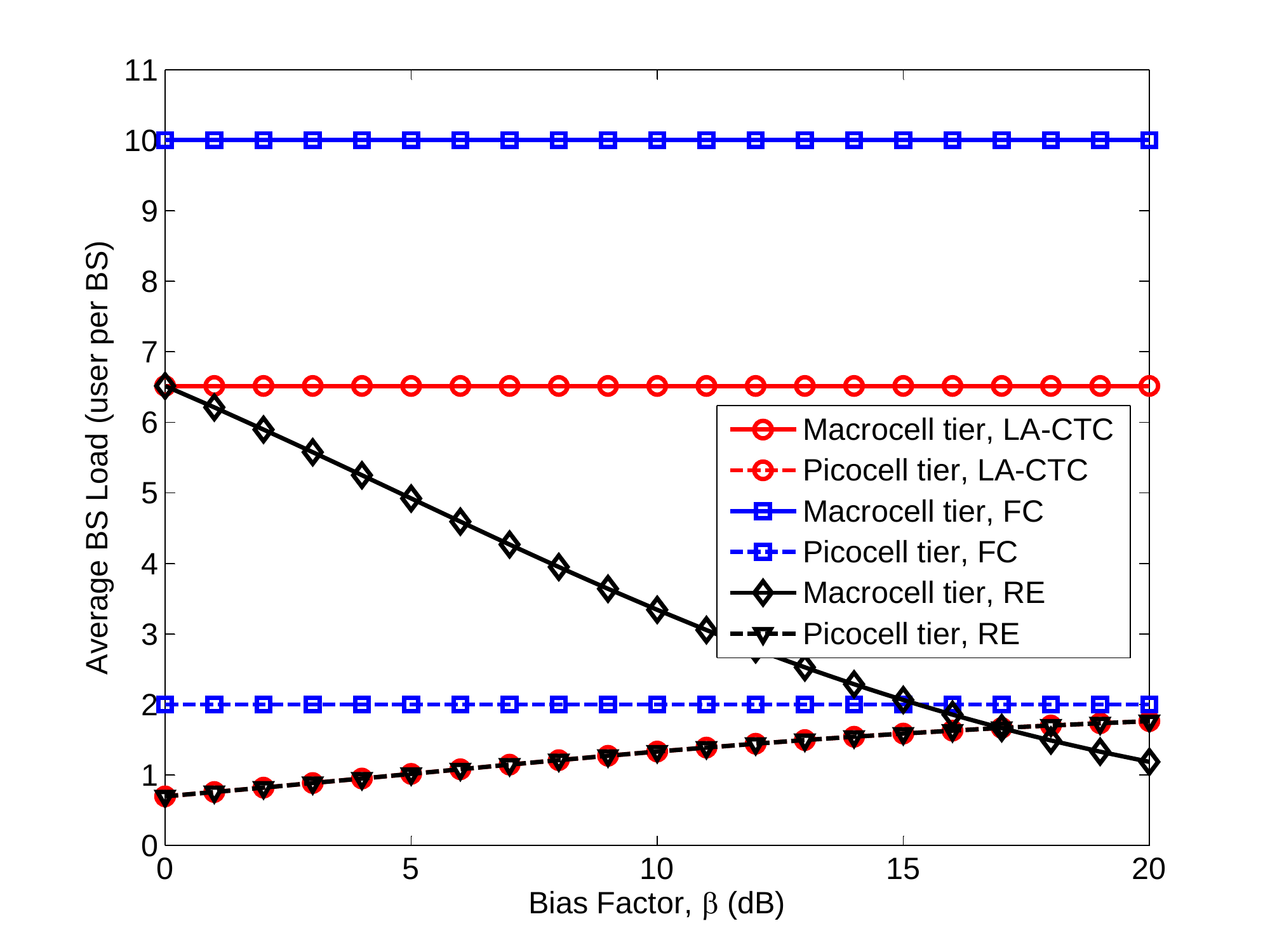}
\caption{LA-CTC vs. Range Expansion and Full Cooperation: Average load per BS vs. the cooperation threshold (bias factor) $\beta$ where $\lambda_1 = (500^2 \pi)^{-1}$, $\lambda_2 = 5 (500^2 \pi)^{-1}$, $\lambda_u = 10 (500^2 \pi)^{-1}$, $P_1 = 37$ dBm, $P_2 = 20$ dBm, $\alpha_1 = \alpha_2 = 3.5$, and $\sigma_z^2 = -104$ dBm.}
\label{load_vs_beta}
\end{figure}

Fig. \ref{load_vs_beta} shows the impact of increasing the cooperation threshold (bias factor) on the average load per BS for the RE scheme, FC scheme, as well as the LA-CTC scheme. It can be seen that, as the bias factor increases, the RE scheme reduces the number of users per macro BS compared to the Tr scheme without biasing (i.e, when $\beta = 0$ dB), by offloading some of the macro users to the pico-tier based on the received powers at these users. On the other hand, the FC scheme increases the number of users per both macro BS and pico BS compared to the Tr  scheme in a two-tier cellular network since it serves all users by using cooperation between BSs in the two tiers. Finally, it can be seen that the LA-CTC scheme keeps the same number of users per macro BS while increasing the number of users per pico BS when compared to the Tr scheme. This is due to the fact that the proposed scheme does not actually offload any users to a different tier. Instead, it changes the mode of operation of users with bad ${\rm SINR}$ conditions which are now served by the original macro BS along with the strongest interfering pico BS. 

The load per BS can reflect the amount of backhaul data exchange required by each scheme. For example, none of the Tr and RE schemes  requires any users' data exchange between any two BSs over the backhaul links since all users are served by a single BS all the time. On the other hand, among the four schemes,  the FC scheme requires the maximum amount of backhaul data exchange  since it uses cooperation to serve all users. In our proposed scheme, the amount of backhaul data exchange lies between those of the Tr and RE, and FC schemes.

\begin{figure*}[!t]
\centering
\subfloat[Case I: Full Cooperation scheme, $\alpha_1 = \alpha_2 = 4$]{\includegraphics[width=0.3\textwidth]{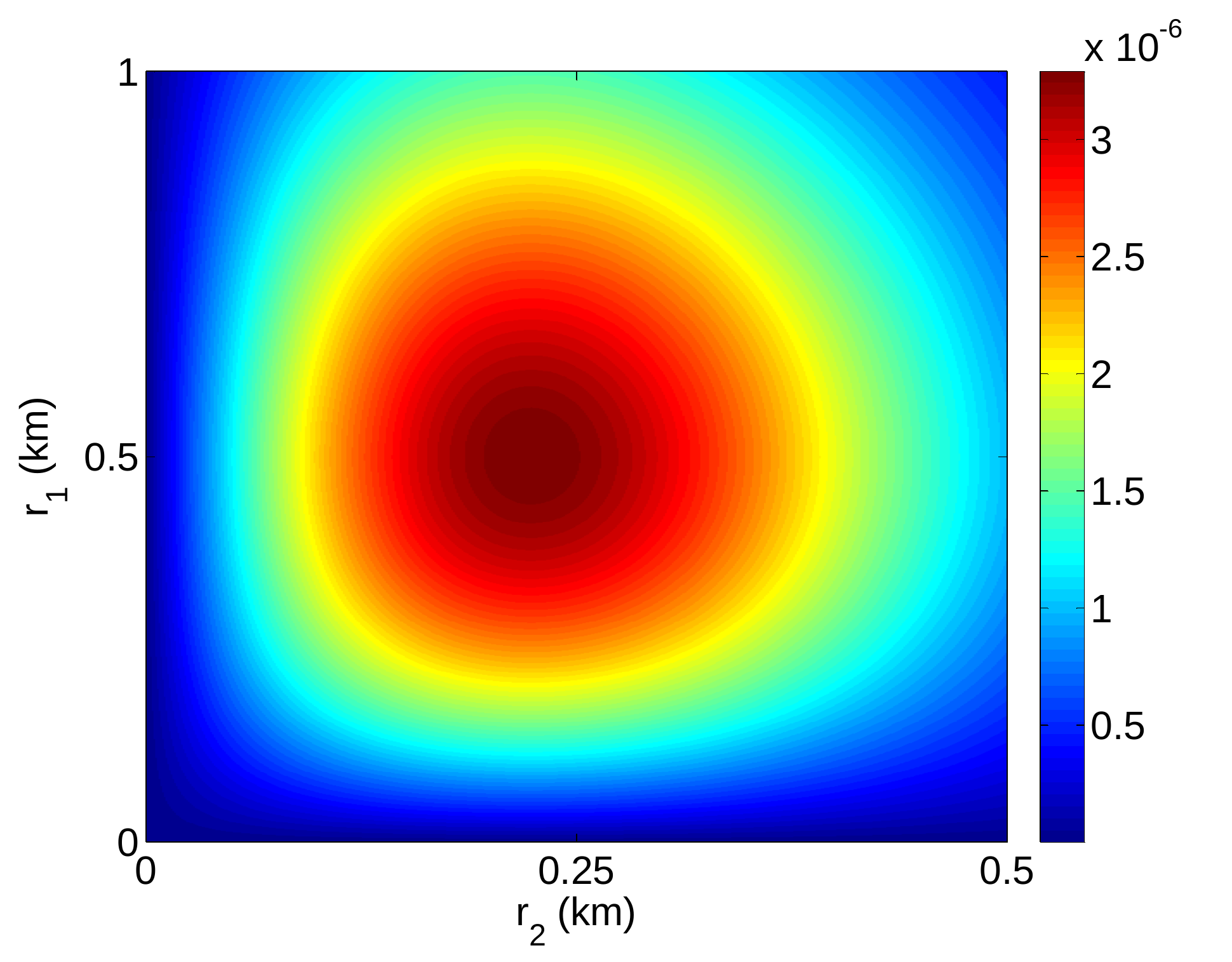}
\label{fig_first_case}}
\hfil
\subfloat[Case II: LA-CTC scheme, $\alpha_1 = \alpha_2 = 4$ and $\beta = 4$ dB ]{\includegraphics[width=0.3\textwidth]{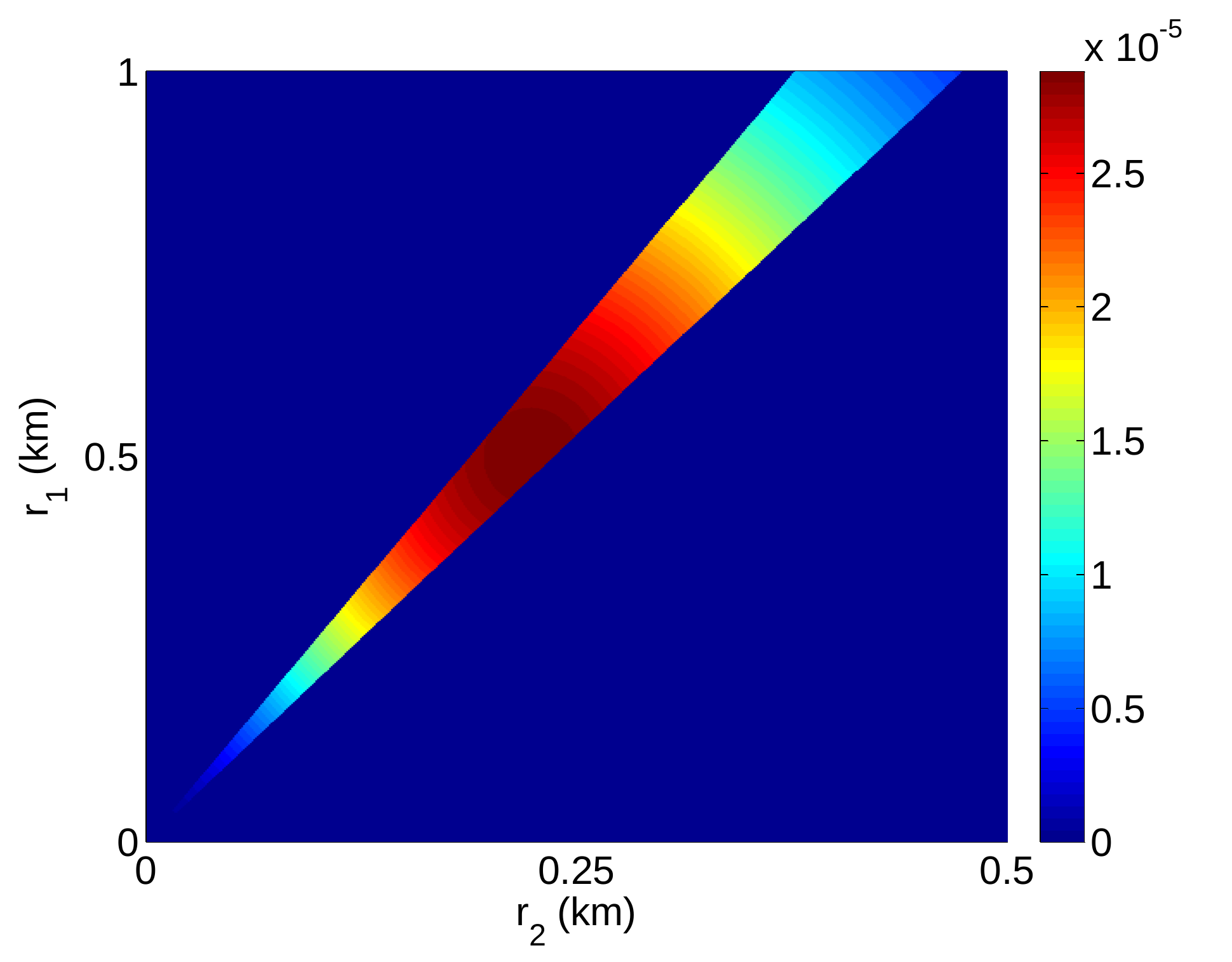}
\label{fig_second_case}}
\\
\subfloat[Case III: LA-CTC scheme, $\alpha_1 = \alpha_2 = 4$ and $\beta = 10$ dB]{\includegraphics[width=0.3\textwidth]{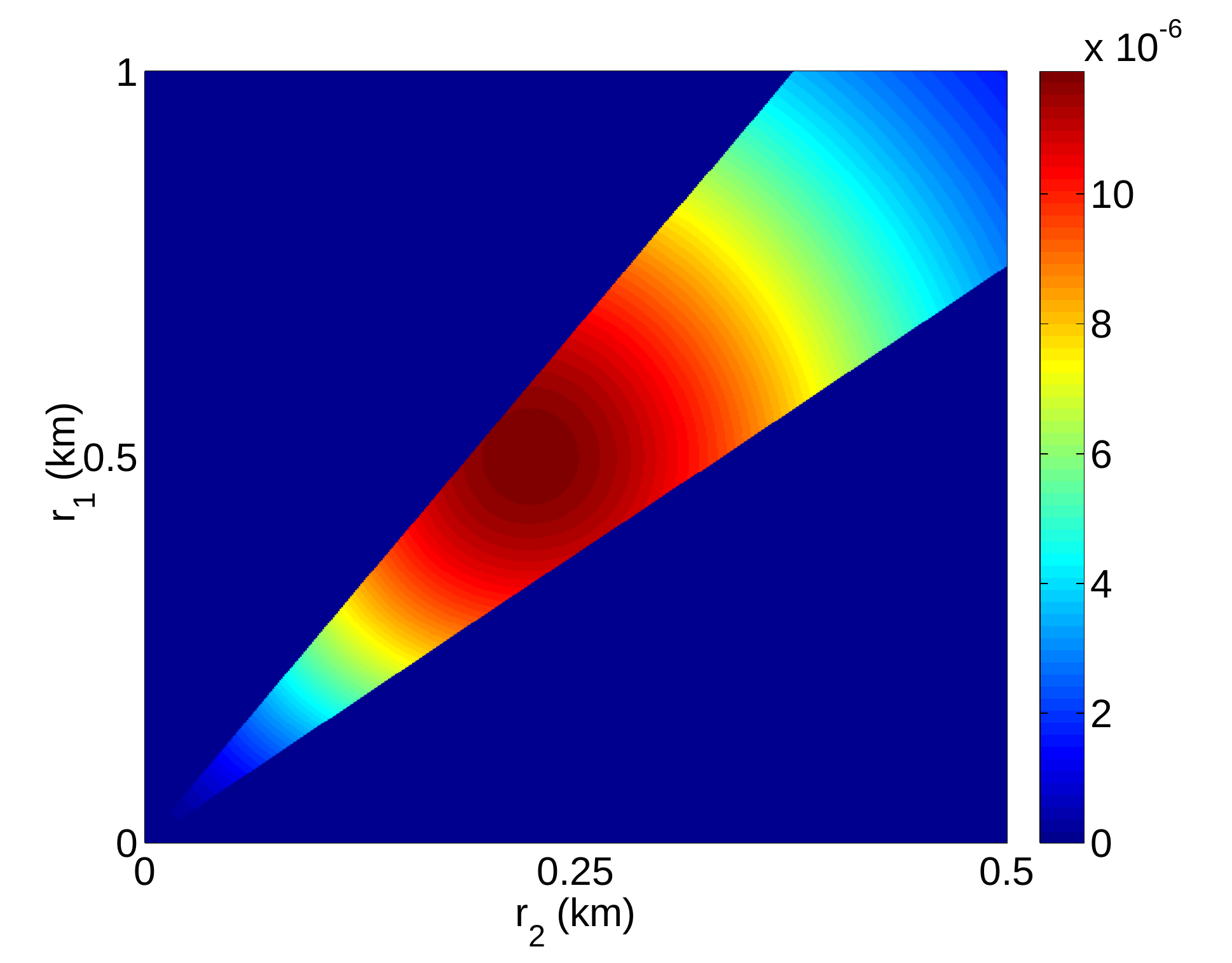}
\label{fig_third_case}}
\hfil
\subfloat[Case IV: LA-CTC scheme, $\alpha_1 =3.5$, $\alpha_2 = 4$ and $\beta = 4$ dB]{\includegraphics[width=0.3\textwidth]{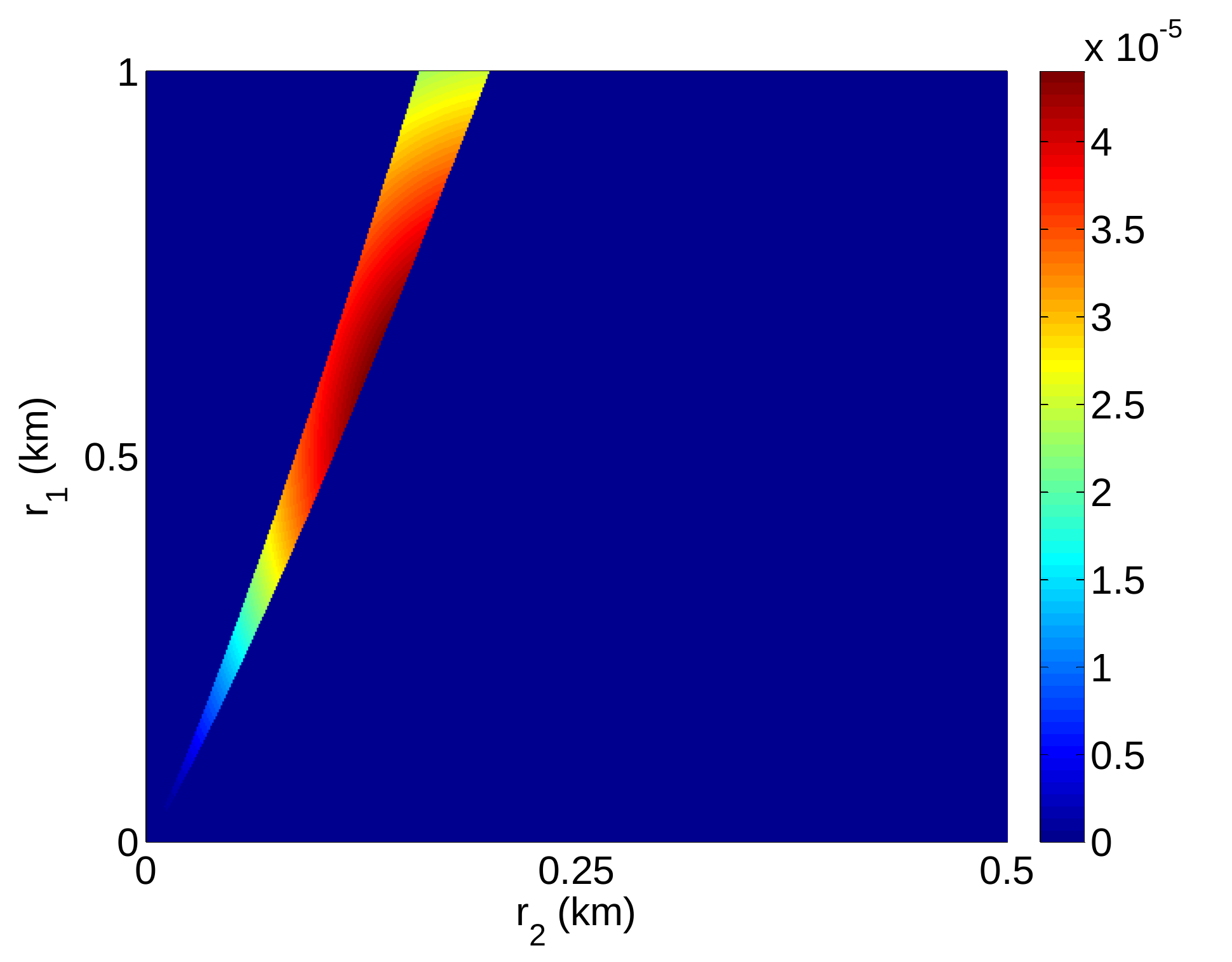}
\label{fig_fourth_case}}
\vspace{3mm}
\caption{LA-CTC vs. Full Cooperation: Joint PDF of the distance of a CoMP user to serving BSs where $\lambda_1 = (500^2 \pi)^{-1}$, $\lambda_2 = 5 (500^2 \pi)^{-1}$, and $P_1 = 37$ dBm, $P_2 = 20$ dBm.}
\label{joint_pdf}
\end{figure*}

In order to compare the FC scheme with the proposed scheme, Fig. \ref{joint_pdf} shows the joint PDF of the distance of a CoMP user to serving BSs for both the schemes. It can be seen in Fig. \ref{fig_first_case} that the FC scheme serves all users by CoMP transmission regardless of their locations in the network. For the LA-CTC scheme, Figs. \ref{fig_second_case} and \ref{fig_third_case} show the effect of increasing the cooperation threshold on the area of cooperation region. With a higher cooperation threshold $\beta$, more users are included in the cooperation regions which, in turn, increases the amount of users' data exchange over the backhaul network. Compared to Fig. \ref{fig_first_case}, it can be seen that users with good ${\rm SINR}$ conditions, who are close to the serving BS and far from the strongest interferer, do not use CoMP transmission to save the resources of the backhaul network. Fig. \ref{fig_fourth_case} shows that the effect of cross-tier interference decreases when the path-loss exponent of the pico-tier is higher than that of the macro-tier, which isolates the pico cells. That is, CoMP transmission is limited to users who are very close to the pico BSs and thus the amount of required data exchanges is reduced.

\section{Conclusion}
\label{sec:conc}
We have investigated the concept of cross-tier cooperation in two-tier cellular networks. We have proposed a novel location-aware cross-tier cooperation scheme that uses downlink CoMP transmission depending on the locations of the users and their nearest macro and pico BSs. Tools from stochastic geometry have been used to analyze the outage probability and average rate for the proposed scheme. The proposed scheme has been compared with three other schemes, namely, Traditional (Tr), Range Expansion (RE), and Full Cooperation (FC) schemes. The comparison has been performed in terms of outage probability, average ergodic rate, as well as load per BS. The results have shown that the proposed LA-CTC scheme outperforms both the range expansion and traditional schemes in terms of outage probability and average ergodic rate. However, this performance gain with the proposed scheme comes at the expense of the overhead due to the exchange of users' data between the two different BSs. In addition, the performance of the proposed scheme approaches that of the FC scheme for sufficiently high cooperation threshold. In this way, the LA-CTC scheme provides a tradeoff between the improved outage probability and the cost of cooperation between BSs in terms of load per BS which reflects the amount of users' data exchange over the backhaul network. As a future extension to this work, cooperation between co-tier BSs could be exploited  to mitigate the effect of co-tier interference as well. In addition, further work is needed to take into account the effect of non-ideal backhaul links on the performance gain of the proposed scheme.

\appendices
\renewcommand{\thesubsection}{\thesection -\Roman{subsection}}
\def\thesubsectiondis{\Roman{subsection}.} 
\section{Proof of Lemma \ref{lem1}}
\label{lem1proof}

Firstly, we derive the probability for a typical user to operate in a certain mode. By definition,
\begin{align}
q_M &= \mathbb{E}_{R_1} \left[ \mathbb{P} \left[ \mathcal{B}=x_1 \right]  \right] = \mathbb{E}_{R_1} \left[ \mathbb{P} \left[ \frac{P_1 R_1^{-\alpha_1}}{P_2 R_2^{-\alpha_2}} \geq \beta \right]  \right], \label{eq:q_M_def}
\end{align}
and
\begin{align}
q_P &= \mathbb{E}_{R_2} \left[ \mathbb{P} \left[ \mathcal{B}=x_2 \right]  \right] = \mathbb{E}_{R_2} \left[ \mathbb{P} \left[ \frac{P_1 R_1^{-\alpha_1}}{P_2 R_2^{-\alpha_2}} < 1 \right]  \right].\label{eq:q_P_def}
\end{align}
Using (\ref{eq:q_M_def}) and (\ref{eq:q_P_def}) and following the proof of Lemma $1$ in \cite{6287527} with the proper changes, $q_M$, $q_P$, and $q_C$ can be obtained.

For the joint PDF $f_{R_C}(\mathbf{r})$ of a typical CoMP user's distance to the cooperating macro BS and pico BS, we know for sure that if the distance to the macro BS is $R_1$, the distance to the pico BS $R_2$ is bounded as follows:
\begin{align}
\left(\frac{P_2}{P_1}\right)^{\frac{1}{\alpha_2}} R_1^{\frac{\alpha_1}{\alpha_2}} < R_2 < \left(\frac{ \beta P_2}{P_1}\right)^{\frac{1}{\alpha_2}} R_1^{\frac{\alpha_1}{\alpha_2}}, \label{eq:bound}
\end{align}
as can be obtained from (\ref{eq:coop_set}) when $\mathcal{B} = \{ x_1, x_2\}$.

Therefore, the conditional probability of $R_1 > r_1$ and $R_2 > r_2$ given that the user operates in the CoMP mode can be written as
\begin{align}
&\mathbb{P}\left[ R_1 > r_1, R_2 > r_2 | \mathcal{B} = \{x_1,x_2\} \right] \nonumber\\
				&\stackrel{(a)}{=} \frac{1}{q_C}\mathbb{P}\left[ R_1 > r_1, R_2 > \max \left( r_2,  \left(\frac{P_2}{P_1}\right)^{\frac{1}{\alpha_2}} R_1^{\frac{\alpha_1}{\alpha_2}}\right)\right]   \nonumber\\
		&-\frac{1}{q_C}\mathbb{P}\left[ R_1 > r_1, R_2 > \max \left(r_2, \left(\frac{ \beta P_2}{P_1}\right)^{\frac{1}{\alpha_2}} R_1^{\frac{\alpha_1}{\alpha_2}}\right) \right]  \nonumber\\
		&\stackrel{(b)}{=} \frac{1}{q_C} \int \limits_{r>r_1} \left( \mathbb{P}\left[ R_2 > \max \left( r_2,  \left(\frac{P_2}{P_1}\right)^{\frac{1}{\alpha_2}} R_1^{\frac{\alpha_1}{\alpha_2}}\right) \right] \right. \nonumber \\
	&~~~~~~~~~\left. - \mathbb{P}\left[ R_2 > \max \left(r_2, \left(\frac{ \beta P_2}{P_1}\right)^{\frac{1}{\alpha_2}} R_1^{\frac{\alpha_1}{\alpha_2}}\right) \right] \right) f_{R_1}'(r) \text{d}r \label{eq:distance_cdf}
\end{align}
where (a) follows the bound on $R_2$ given in (\ref{eq:bound}) and $f_{R_i}'(r)$ in (b) is the distribution of the distance to the nearest point in a homogeneous PPP $\Phi_i \in \mathbb{R}^2$ which can be derived as follows:
\begin{align}
\mathbb{P}\left[ R_i > r \right] &= \mathbb{P}\left[ \text{There are no BSs in a disc of radius $r$} \right] \nonumber\\
											& = \exp \left[ -\pi \lambda_i r^2 \right].
\end{align}
Therefore,
\begin{align}
f_{R_i}'(r) &= \frac{\text{d}}{\text{d} r} \left(1 - \mathbb{P}\left[ R_i > r \right] \right)\nonumber \\ 
															&= 2 \pi \lambda_i ~r \exp \left[ -\pi \lambda_i r^2 \right] \label{eq:distance_dist}.
\end{align}

After plugging (\ref{eq:distance_dist}) into (\ref{eq:distance_cdf}), we use the resulting cumulative CDF (CCDF), i.e., $\mathbb{P}\left[ R_1 > r_1, R_2 > r_2 | \mathcal{B} = \{x_1,x_2\}\right]$, to obtain the joint PDF $f_{R_C}(\mathbf{r})$ of $R_1$ and $R_2$ of a user who operates in the CoMP mode as follows:
\begin{align}
&f_{R_C}(\mathbf{r}) = \frac{\partial^2}{\partial r_1 \partial r_2} \left( 1-  \mathbb{P}\left[ R_1 > r_1, R_2 > r_2 | \mathcal{B} = \{x_1,x_2\}\right]\right) \nonumber\\
			&= \left\{ 
			\begin{array} {l l}
				\frac{4 \pi^2 \lambda_1 \lambda_2}{q_C} r_1 r_2 \exp \left[ -\pi \left( \lambda_1 r_1^2 + \lambda_2 r_2^2 \right) \right], & (r_1,r_2)\in \mathcal{A}\\
				0, & \text{otherwise}
			\end{array}
			\right.
\end{align}
where 
\begin{align}
\mathcal{A} = \Scale[0.90]{\left\{ (r_1,r_2) : r_1 \geq 0 \text{ and } \left(\frac{P_2}{P_1}\right)^{\frac{1}{\alpha_2}} r_1^{\frac{\alpha_1}{\alpha_2}} < r_2 < \left(\frac{ \beta P_2}{P_1}\right)^{\frac{1}{\alpha_2}} r_1^{\frac{\alpha_1}{\alpha_2}} \right\}}. \label{eq:A}
\end{align}

For $f_{R_1}(r)$, we use the event of $R_1 > r$ given that the macro user operates in the non-CoMP mode, i.e., $\mathcal{B} = \{x_1\}$, where the probability of this event is given by
\begin{align}
\mathbb{P}[ R_1 > r_1 &| \mathcal{B} = \{x_1\} ] = \frac{1}{q_M}\mathbb{P}\left[ R_1 > r_1, \frac{P_1 R_1^{-\alpha_1}}{P_2 R_2^{-\alpha_2}} > \beta \right]  \nonumber\\
		&= \frac{1}{q_M} \int \limits_{r>r_1} \mathbb{P}\left[ R_2 > \left(\frac{\beta P_2}{P_1}\right)^{\frac{1}{\alpha_2}} r^{\frac{\alpha_1}{\alpha_2}} \right]  f_{R_1}'(r) \text{d}r. \label{eq:distance_cdf_1}
\end{align}

Then, we follow the same procedure by plugging (\ref{eq:distance_dist}) into (\ref{eq:distance_cdf_1}) and taking the the derivative of the CDF, i.e., $1- \mathbb{P}\left[ R_1 > r_1 | \mathcal{B} = \{x_1\} \right]$, with respect to $r_1$. Hence,  (\ref{eq:distance_cdf_1}) reduces to (\ref{eq:dist_M}). Similarly, we can obtain the PDF of $R_2$ as in (\ref{eq:dist_P}).

\section{Proof of Theorem \ref{thm1}}
\label{thm1proof}

Firstly, we derive the outage probability of a randomly located non-CoMP macro user. Using the definition of the outage probability in (\ref{eq:outage_gen}) for a non-CoMP macro user,
\begin{align}
\mathcal{O}_M &= 1 - \int \limits_{\mathbb{R}_+} \mathbb{P} \left[ \rm{SINR}(\mathcal{B}=\{x_1\}) > \tau \right] f_{R_1}(r_1) \text{d}r_1 \label{eq:O_M_def}
\end{align}
where the ${\rm SINR}$ in (\ref{eq:SINR}) can be rewritten as
\begin{align}
\rm{SINR}(\mathcal{B}=\{x_1\}) &= \frac{P_1 |h_{1,0}|^2 r_1^{-\alpha_1}}{I + \sigma_z^2}
\end{align}
in which
\begin{align}
I &= I_1 + I_2,\nonumber\\
I_1 &= P_1 \sum_{x_i \in \Phi_1 \setminus x_1} |g_{1,i}|^2 \|x_i\|^{-\alpha_1} \nonumber\\
I_2 &= P_2 \sum_{x_i \in \Phi_2} |g_{2,i}|^2 \|x_i\|^{-\alpha_2}.
\end{align}
$I_i$ ($i \in \{1,2\}$) is the total interference power received from the $i^{th}$ tier and $f_{R_1}(r_1)$ is the PDF of distance given in \textbf{Lemma \ref{lem1}}.

After rewriting the ${\rm SINR}$ of the non-CoMP macro user, we can calculate the CCDF as follows:
\begin{align}
\mathbb{P} \left[ \rm{SINR} > \tau \right] &= \mathbb{P} \left[ |h_{1,0}|^2 > \tau \frac{I + \sigma_z^2}{P_1 r_1^{-\alpha_1}} \right] \nonumber\\
	&\stackrel{(c)}{=} \int \limits_{\mathbb{R}_+} \exp \left[ -\tau \frac{i + \sigma_z^2}{P_1 r_1^{-\alpha_1}} \right] f_I(i) di \nonumber\\
	&= \mathbb{E}_I \left[ \exp \left[ -\tau \frac{i + \sigma_z^2}{P_1 r_1^{-\alpha_1}} \right] \right] \nonumber\\
	&\stackrel{(d)}{=} \exp \left[\frac{- \tau \sigma_z^2}{P_1 r_1^{-\alpha_1}} \right] \prod_{j=1}^2 \mathcal{L}_{I_j}\left( \frac{\tau r_1^{\alpha_1}}{P_1 } \right)  \label{eq:SINR_CCDF}
\end{align}
where (c) follows because the channel fading power $|h_{1,0}|^2 \sim \Exp(1)$, and (d) follows from the definition of Laplace transform. Without loss of generality, we calculate the Laplace transform of $I_1$ and the calculation of the Laplace transform of $I_2$ follows the same procedure.
\begin{align}
\mathcal{L}_{I_1}\left( s \right) &= \mathbb{E}_I \left[ \exp \left[ -s I_1 \right] \right] \nonumber\\
	&= \mathbb{E}_{\Phi_1,\{g_{1,i}\}} \left[ \exp \left[ -s P_1 \sum_{x_i \in \Phi_1 \setminus x_1} |g_{1,i}|^2 R_i^{-\alpha_1} \right] \right] \nonumber\\
	&\stackrel{(e)}{=} \mathbb{E}_{\Phi_1} \left[\prod_{x_i \in \Phi_1 \setminus x_1} \mathbb{E}_{\{g_{1,i}\}} \left[ \exp \left[ -s P_1  |g_{1,i}|^2 R_i^{-\alpha_1}\right] \right] \right] \nonumber\\
	&\stackrel{(f)}{=} \mathbb{E}_{\Phi_1} \left[\prod_{x_i \in \Phi_1 \setminus x_1} \frac{1}{1+s P_1 R_i^{-\alpha_1}} \right]\nonumber \\
	&\stackrel{(g)}{=} \exp \left[ -2 \pi \lambda_1 \int \limits_{r>r_1} \left( 1 - \frac{1}{1+s P_1 r^{-\alpha_1}} \right) r \text{d}r \right]. \label{eq:Laplace}
\end{align}
In the above, (e) follows because of the independence assumption between $g_{1,i}$'s, (f) follows because the moment generating function of an exponential random variable with parameter $\mu$ is $\mu/(\mu-t)$, while (g) follows the probability generating functional of PPP. Now, let $u^{\alpha_1} = (s P_1)^{-1} r^{\alpha_1}$ and replacing $s$ with $\frac{\tau r_1^{\alpha_1}}{P_1}$, we obtain
\begin{align}
\Scale[1]{\mathcal{L}_{I_1}\left( \frac{\tau r_1^{\alpha_1}}{P_1} \right)} &= \Scale[1]{\exp \left[ - 2 \pi \lambda_1 \tau^{\frac{2}{\alpha_1}} r_1^2 \mathcal{F}\left((\frac{1}{\tau})^{\frac{1}{\alpha_1}},\alpha_1\right)\right]} \label{eq:L_1}
\end{align}
where $\mathcal{F}\left(y,\alpha\right)$ is defined in (\ref{eq:F}). Similarly, we can obtain the Laplace transform of $I_2$ as
\begin{align}
\Scale[1]{\mathcal{L}_{I_2}\left( \frac{\tau r_1^{\alpha_1}}{P_1 } \right)} &= \Scale[1]{\exp \left[ - 2 \pi \lambda_2  \left(\tau \frac{P_2}{P_1}\right)^{\frac{2}{\alpha_2}} r_1^{\frac{2\alpha_1}{\alpha_2}} \mathcal{F}\left((\frac{\beta}{\tau})^{\frac{1}{\alpha_1}} ,\alpha_2\right)\right]} \label{eq:L_2}
\end{align}
where the closest interferer in this case is at least at a distance $ \left(\frac{\beta P_2}{P_1}\right)^{\frac{1}{\alpha_2}} r_1^{\frac{\alpha_1}{\alpha_2}}$ instead of $r_1$ which was used to obtain the Laplace transform of $I_1$. By combining (\ref{eq:L_1}) and (\ref{eq:L_2}) with (\ref{eq:SINR_CCDF}) and then substituting in (\ref{eq:O_M_def}), we obtain (\ref{eq:O_M}). The outage probability of a non-CoMP pico user can be easily obtained as in (\ref{eq:O_P}) by following the same procedure.

For a randomly located CoMP user, given that $\mathcal{B} = \{x_1,x_2\}$, the outage probability is given by
\begin{align}
\mathcal{O}_C &= 1 - \int \limits_{\mathcal{A}} \mathbb{P} \left[ \rm{SINR}(\mathcal{B} = \{x_1,x_2\}) > \tau \right] f_{R_c}(\mathbf{r}) \text{d}\mathbf{r} \label{eq:O_C_def}
\end{align}
where $f_{R_c}(\mathbf{r})$ is the joint PDF of the distance to the nearest two BSs (one from each tier) to the typical user, i.e., $x_1$ and $x_2$, given in \textbf{Lemma \ref{lem1}}. Then, we can rewrite the ${\rm SINR}$ in (\ref{eq:SINR}) as
\begin{align}
\rm{SINR}(\mathcal{B}) &= \frac{|\sqrt{P_1} h_{1,0} r_1^{-\frac{\alpha_1}{2}} + \sqrt{P_2} h_{2,0} r_2^{-\frac{\alpha_2}{2}}|^2}{ I + \sigma_z^2}
\end{align}
where
\begin{align}
I &= I_1 + I_2,\nonumber\\
I_1 &= P_1 \sum_{x_i \in \Phi_1 \setminus x_1} |g_{1,i}|^2 \|x_i\|^{-\alpha_1} \nonumber\\
I_2 &= P_2 \sum_{x_i \in \Phi_2 \setminus x_2} |g_{2,i}|^2 \|x_i\|^{-\alpha_2}.
\end{align}
Before calculating the CCDF of the ${\rm SINR}$, we define a new variable $\theta_i$ such that
\begin{align}
\theta_i &= \sqrt{P_i} \|x_i\|^{-\frac{\alpha_i}{2}}. \nonumber
\end{align}
Then, the CCDF of the ${\rm SINR}$ can be rewritten as
\begin{align}
\mathbb{P} \left[ \rm{SINR} > \tau \right] &= \mathbb{P} \left[ |\theta_1 h_{1,0} + \theta_2 h_{2,0}|^2 > \tau (I + \sigma_z^2) \right].
\end{align}
Since $h_{i,0}$s are i.i.d. and $\sim \mathcal{CN}(0,1)$, we obtain
\begin{align}
|\theta_1 h_{1,0} + \theta_2 h_{2,0}|^2 \sim \Exp \left( \frac{1}{\sum_{i=1}^2 \theta_i^2} \right) \nonumber
\end{align}
which means that the CCDF of the ${\rm SINR}$ can be written as
\begin{align}
\mathbb{P} \left[ \rm{SINR} > \tau \right] &= \mathbb{E}_I \left[ \exp \left[ -\tau \frac{i + \sigma_z^2}{\sum_{i=1}^2 \theta_i^2} \right] \right] \nonumber\\
	&\stackrel{(g)}{=} \exp \left[\frac{- \tau \sigma_z^2}{\sum_{i=1}^2 \theta_i^2} \right] \prod_{j=1}^2 \mathcal{L}_{I_j}\left( \frac{\tau}{\sum_{i=1}^2 \theta_i^2} \right) \label{eq:SINR_C}
\end{align}
where (g) follows the definition of the Laplace transform of $I_j$. By following the same steps  in deriving  (\ref{eq:Laplace}), we have
\begin{align}
\mathcal{L}_{I_j}\left( s \right) &= \exp \left[ - 2 \pi \lambda_j (s P_j)^{\frac{2}{\alpha_j}} \mathcal{F}\left((\frac{1}{s P_j})^{\frac{1}{\alpha_j}} r_j,\alpha_j\right)\right]. \label{eq:L_C}
\end{align}
By combining (\ref{eq:dist_C}), (\ref{eq:SINR_C}), and (\ref{eq:L_C}), and then substituting in (\ref{eq:O_C_def}), we obtain the outage probability of a randomly located CoMP user as given in (\ref{eq:O_C}).

\section{$\mathcal{F}(y,\alpha)$ Special Cases}
\label{F_fun}

The function $\mathcal{F}$, given in (\ref{eq:F}), has a semi-open integral and does not give a closed-form solution in general. However, this function yields a closed-form expression for some values of $\alpha$. For example, if $\alpha$ is a rational number and can be expressed as
\begin{align}
\alpha &= \frac{2n}{n-m}, \qquad n>m \nonumber
\end{align}
where $n$ and $m$ are any two positive integer numbers, the function $\mathcal{F}$ reduces to
\begin{align}
\mathcal{F}(y,\alpha) &= \frac{(-1)^{\frac{2}{\alpha}}}{\alpha}  \sum_{k=0}^{n-1} \frac{\ln\left[ 1-\sqrt[n]{-y^{-\alpha}} \exp\left[\frac{2\pi i k}{n}\right] \right] }{\exp\left[ \frac{2\pi i k (\alpha-2)}{\alpha } \right]}
\end{align}
where $i = \sqrt{-1}$ is the imaginary unit number.

This expression reduces to even more simpler expressions for specific values of $\alpha$. For example, if $\alpha = 4$, i.e., $m=1$ and $n=2$, $\mathcal{F}(y^{\frac{-1}{4}},4)$ reduces to
\begin{align}
\mathcal{F}(y^{\frac{-1}{\alpha}},\alpha) &= \frac{1}{2} \arctan(\sqrt{y}). \nonumber
\end{align}

\begin{IEEEbiography} [{\includegraphics[width=1in,height=1.25in,clip,keepaspectratio]{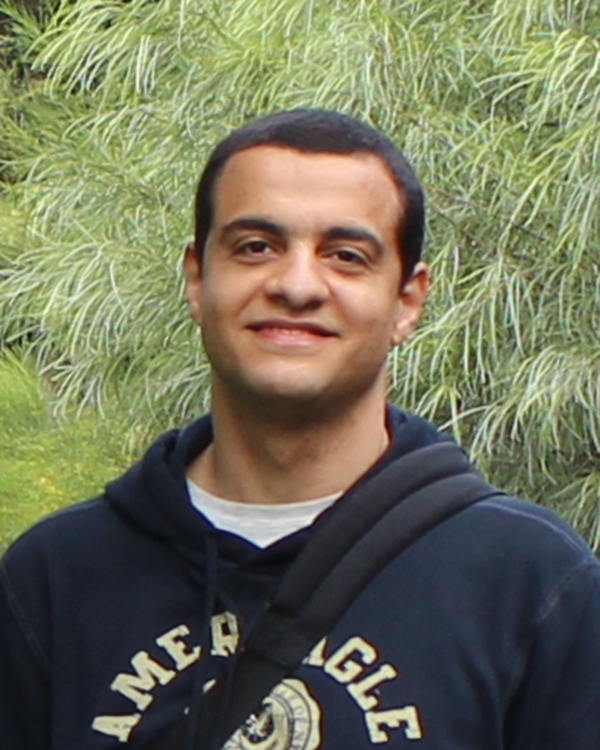}}]
{Ahmed H. Sakr} (S'12) is a Ph.D. candidate in the Department of Electrical and Computer Engineering, University of Manitoba, Canada.  He received the B.Sc. (2002-2007) and M.Sc. (2010-2012) degrees both in Electronics and Communications Engineering from Tanta University, Tanta, Egypt, and Egypt-Japan University of Science and Technology (E-JUST), Alexandria, Egypt, respectively. For his academic excellence, he has received several academic awards including the University of Manitoba Graduate Fellowship (UMGF) in 2014-2016, the Graduate Enhancement of Tri-Council Stipends (GETS) in 2013, and Egyptian Ministry of Higher Education Excellence Scholarship in 2010-2012. Ahmed has been a member in the technical program committee and a reviewer in several IEEE journals and conferences. His current research interests include statistical modeling of wireless networks, resource allocation in multi-tier cellular networks, and green communications.
\end{IEEEbiography}

\begin{IEEEbiography} [{\includegraphics[width=1in,height=1.25in,clip,keepaspectratio]{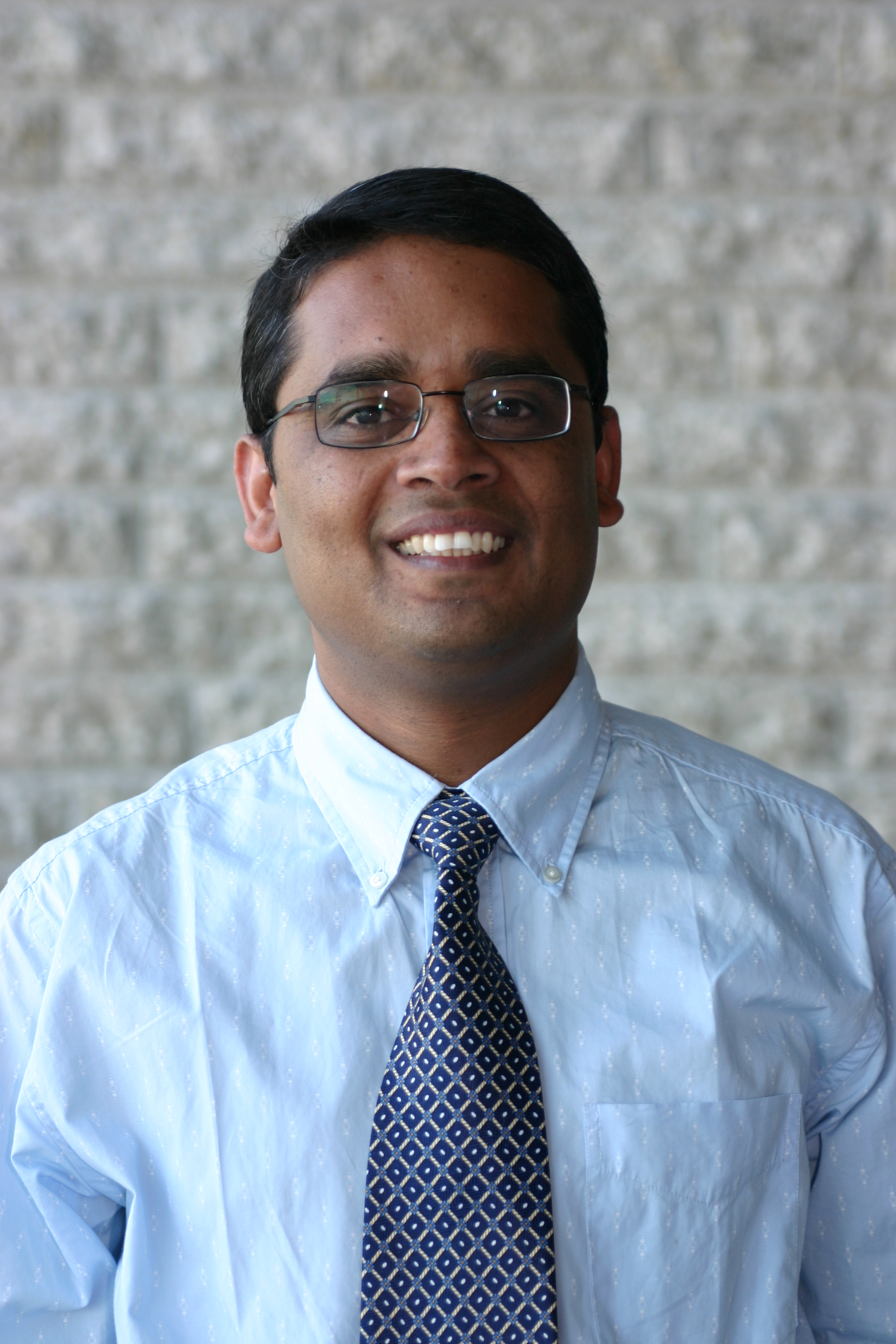}}]
{Ekram Hossain} (S'98-M'01-SM'06)
is a Professor (since March 2010) in the Department of Electrical and Computer Engineering at University of Manitoba, Winnipeg, Canada. He received his Ph.D. in Electrical Engineering from University of Victoria, Canada, in 2001. Dr. Hossain's current research interests include design, analysis, and optimization of wireless/mobile communications networks, cognitive radio systems, and network economics. He has authored/edited several books in these areas (http://home.cc.umanitoba.ca/$\sim$hossaina). Dr. Hossain serves as the Editor-in-Chief for the {\em IEEE Communications Surveys and Tutorials} and an Editor for {\em IEEE Journal on Selected Areas in Communications - Cognitive Radio Series} and {\em IEEE Wireless Communications}. Also, he is a member of the IEEE Press Editorial Board. Previously, he served as the Area Editor for the {\em IEEE Transactions on Wireless Communications} in the area of ``Resource Management and Multiple Access'' from 2009-2011 and an Editor for the IEEE Transactions on Mobile Computing from 2007-2012. Dr. Hossain has won several research awards including the University of Manitoba Merit Award in 2010 and 2014 (for Research and Scholarly Activities), the 2011 IEEE Communications Society Fred Ellersick Prize Paper Award, and the IEEE Wireless Communications and Networking Conference 2012 (WCNC'12) Best Paper Award. He is a Distinguished Lecturer of the IEEE Communications Society (2012-2015). Dr. Hossain is a registered Professional Engineer in the province of Manitoba, Canada.
\end{IEEEbiography}

\end{document}